\documentclass[preprint,12pt]{elsarticle}

\usepackage{amssymb}
\usepackage{adjustbox}
\usepackage{eurosym}

\usepackage[T1]{fontenc}
\usepackage{graphicx}
\usepackage{balance}  
\usepackage{xcolor}

\usepackage{enumitem}

\usepackage{ifthen}
\usepackage{hyperref}
\usepackage{stmaryrd}
\usepackage{amsmath}
\usepackage{colortbl}
\usepackage{balance}
\usepackage{marginnote}
\usepackage{multicol}
\usepackage{listings}
\usepackage[linesnumbered]{algorithm2e}
\usepackage{algpseudocode}
\usepackage{mathtools}
\usepackage{bm}
\usepackage{caption}
\usepackage{lipsum}
\usepackage{verbatim}

\usepackage{tikz}
\usepackage{tkz-graph}
\usepackage{pgfplots}
\usepackage{subfig}
\usetikzlibrary{arrows, automata,petri,backgrounds}
\usetikzlibrary{positioning,shapes,shadows,arrows,decorations.markings,snakes}
\usetikzlibrary{decorations.pathmorphing}
\usepackage{wrapfig}

\newcommand{\eat}[1]{}

\newcommand{\edgesin}[2]{N_{\rightarrow {#1}}^{#2}}
\newcommand{\edgesout}[2]{N_{{#1} \rightarrow}^{#2}}

\newtheorem{property}{Property}

\newcommand\mysection[1]{\vspace{-2mm}\section{#1}\vspace{-0.5mm}}
\newcommand\mysubsection[1]{\vspace{-1mm}\subsection{#1}\vspace{-0.5mm}}
\newcommand\va{\vspace{-0mm}}

\newcommand\grow{\textsc{Grow}}
\newcommand\mergecur{\textsc{Merge}}

\newcommand{\extVersion}{false}  
\newcommand{\printIfExtVersion}[2]
{
        \ifthenelse{\equal{\extVersion}{true}}{#1}{}
        \ifthenelse{\equal{\extVersion}{false}}{#2}{}
}

\begin{document}

\begin{frontmatter}

\title{Graph integration of structured, semistructured and unstructured data for data journalism}


\author[ipp]{Angelos~Christos~Anadiotis}
\author[inria]{Oana Balalau}
\author[ist]{Catarina Concei\c{c}\~{a}o}
\author[ist]{Helena Galhardas}
\author[inria]{Mhd~Yamen~Haddad}
\author[inria]{Ioana Manolescu}
\author[inria]{Tayeb Merabti}
\author[inria]{Jingmao You}
\address[ipp]{\'Ecole Polytechnique, Institut Polytechnique de Paris and EPFL}
\address[inria]{Inria, Institut Polytechnique de Paris}
\address[ist]{INESC-ID and IST, Univ. Lisboa, Portugal}

\begin{abstract}
Digital data is a gold mine for modern journalism. However, datasets
which interest journalists
 are extremely heterogeneous, ranging from highly structured
(relational databases), semi-structured (JSON, XML, HTML), graphs
(e.g., RDF), and text. Journalists (and other classes of users lacking
advanced IT expertise, such as most non-governmental-organizations, or
small public administrations) need to
be able to make sense of such heterogeneous corpora, even if they
lack the ability to define and deploy custom extract-transform-load
workflows, especially for dynamically varying sets of data sources.  

We describe a complete approach for integrating dynamic sets of
heterogeneous datasets along the lines described above:  the
challenges we faced to make such graphs useful, allow their integration to scale,
and the solutions we proposed for these problems. Our approach
is implemented within the ConnectionLens system; we validate it
through a set of experiments.

\end{abstract}




\end{frontmatter}

\mysection{Introduction}
\label{sec:intro}

Data journalists often have to analyze and exploit
datasets that they obtain from official organizations or their sources, extract from social media,
or create themselves (typically Excel or Word-style). For instance, journalists from the French Le Monde newspaper want to retrieve {\em connections between elected people at Assembl\'{e}e Nationale and companies that have subsidiaries outside of France}. Such a query can be answered currently at a high human effort cost, by inspecting e.g., a JSON list of Assembl\'{e}e elected officials (available from NosDeputes.fr) and manually connecting the names with those found in a national registry of companies. This considerable effort may still miss connections that could be found if one added information about politicians' and business people's spouses, information sometimes available in public knowledge bases such as DBPedia, or journalists' notes. 

No single query language can be used on such heterogeneous data; instead, we study methods to query the corpus by specifying some keywords and asking for all the connections that exist, in one or across several data sources, between these keywords.
This problem has been studied under the name of {\em keyword search over structured data}, in particular for relational databases~\cite{DBLP:journals/debu/YuQC10,DBLP:conf/vldb/HristidisP02}, XML documents~\cite{guo2003xrank,liu2007identifying}, RDF graphs~\cite{DBLP:journals/tkde/LeLKD14,Elbassuoni:2011:KSO:2063576.2063615}.
 However, most of these works assumed one single source of data, in which connections among nodes are clearly identified. When authors considered several data sources~\cite{Li:2008:EEK:1376616.1376706}, they still assumed that one query answer comes from a single data source.

\begin{figure*}[t!]
\centering
\includegraphics[width=.9\textwidth]{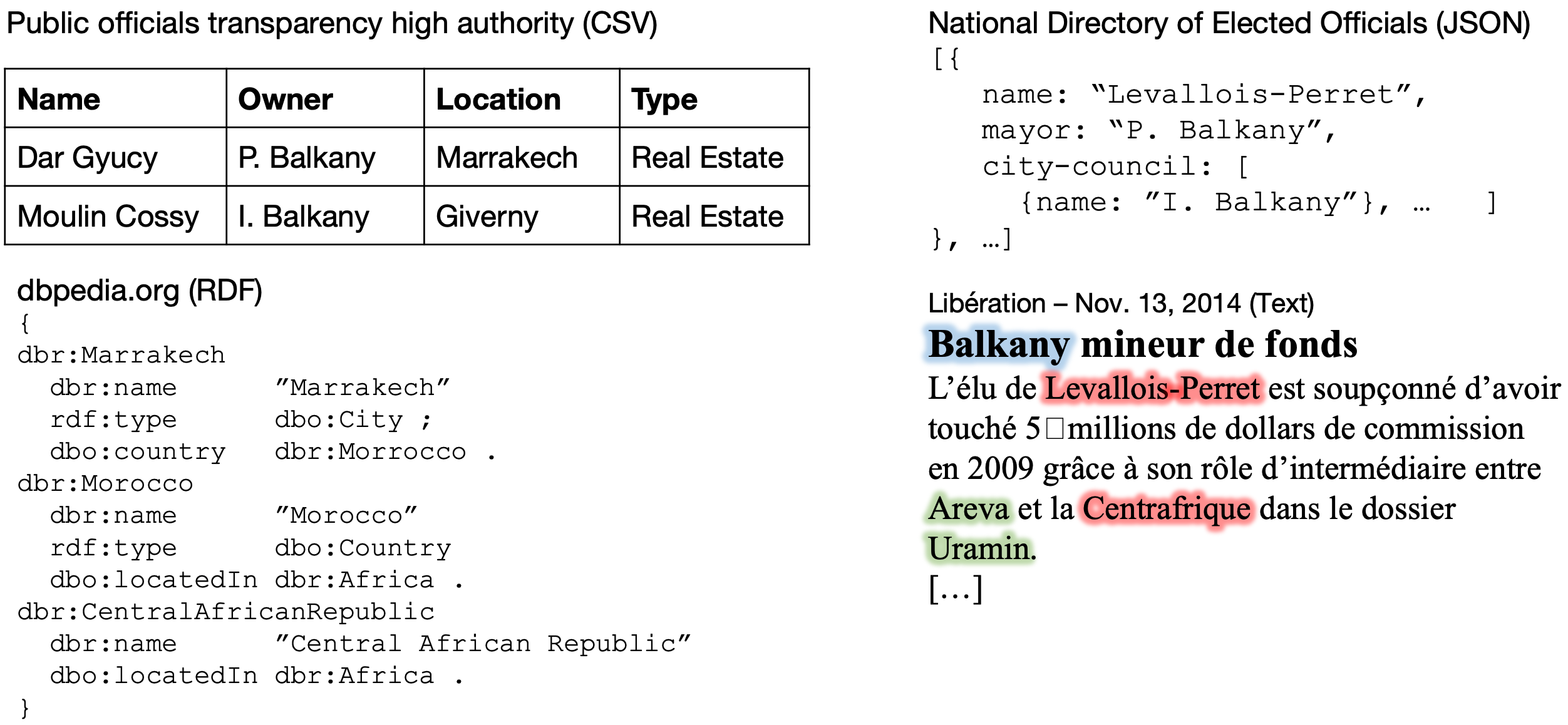}
\vspace{-5mm}
\caption{Motivating example: collection ${\mathcal D}$ of four datasets. \label{fig:motiv-example}}
\vspace{-7mm}
\end{figure*}

In contrast, the ConnectionLens system~\cite{Chanial2018} answers keyword search queries over arbitrary combinations of datasets and heterogeneous data models, independently produced by actors unaware of each other's existence. First, ConnectionLens \textbf{integrates a set of datasets into a unique graph}, subject to the following requirements and constraints:

\textbf{R1. Integral source preservation and provenance:} in journalistic work, it is crucial to be able to trace each node within the integrated graph back to the dataset from which it came. This enables {\em adequately sourcing} information, an important tenet of quality journalism.

\textbf{R2. Little to no effort required from users:} journalists often lack time and resources to set up IT tools or data processing pipelines. Even when they are able to use a tool supporting one or two data models (e.g., most relational databases provide some support for JSON data), handling other data models remains challenging.
Thus, the construction of the integrated graph needs to be as automatic as possible.

\textbf{C1. Little-known entities:} interesting journalistic datasets feature some extremely well-known entities (e.g., highly visible National Assembly members) next to others of much smaller notoriety (e.g., the collaborators employed by the National Assembly to help organize each deputee's work; or a company in which a deputee had worked). From a journalistic perspective, such lesser-known entities may play a crucial role in making interesting connections among nodes in the graph.

\textbf{C2. Controlled dataset ingestion:} the level of confidence in the data required for journalistic use excludes massive ingestion from uncontrolled data sources, e.g., through large-scale Web crawls.

\textbf{C3. Language support:} journalists are first and foremost concerned with the affairs surrounding them (at the local or national scale). This requires supporting dataset in the language(s) relevant for them - in our case, French.

\textbf{R3. Performance on ``off-the-shelf'' hardware:} Our algorithms' complexity in the data size should be low, and overall performance is a concern;  the tool should run on general-purpose hardware, available to non-expert users like the ones we consider.

For what concerns querying the integrated graph, we note:

\textbf{R4. Finding connections across heterogeneous datasets} is a core need. In particular, it is important for our approach to be tolerant of inevitable differences in the organization of data across sources.

\textbf{C4. Query algorithm orthogonal to answer quality scores} After discussing several journalistic scenarios, no unique method (score) for deciding which are the best answers to a query has been identified. Instead: ($i$)~it appears that ``very large'' answers (say, of more than 20 edges) are of limited interest; ($ii$)~connections that ``state the obvious'', e.g., that any two actors from a French political scenario are connected to one another through ``France'', are not of interest. Therefore, our algorithm must be orthogonal and it should be possible to use it with any score function.

To reach our goals under these requirements and constraints, we make the following contributions:

\begin{enumerate}
\va
\item We define the novel {\em integration graphs} we target, and formalize the problem of constructing them from arbitrary sets of datasets.
\va
\item We introduce an  {\em approach}, and an {\em architecture} for building the graphs, leveraging data integration, information extraction,
knowledge bases, and data management techniques. Within this architecture, a significant part of our effort was invested in developing resources and tools for datasets in French. English is also supported, thanks to a (much) wider availability of linguistics resources.
\va
\item We analyze the {\em properties  of the search space} for our keyword search problem, and propose {\em the first algorithm capable of finding matches across heterogeneous data sources} while preserving each node within its original source.
\item We have fully implemented our approach in an end-to-end tool; it currently supports text, CSV, JSON, XML, RDF, PDF datasets, and existing relational databases.  We present: ($i$)~a set of {\em use cases with real journalistic datasets}; ($ii$)~an {\em experimental evaluation} of its scalability and the quality of the extracted graph; ($iii$)~query experiments confirming the practical interest of query algorithm.
\end{enumerate}
\va

\noindent\textbf{Motivating example.} To illustrate our approach, we rely on a set of four datasets, shown in Figure~\ref{fig:motiv-example}.  Starting from the top left, in clockwise order, we have: a table with assets of public officials, a JSON  listing of France elected officials, an article from the newspaper Lib\'{e}ration with entities highlighted, and a subset of the DBPedia RDF knowledge base. Our goal is to {\em interconnect} these datasets into a graph and to be able to answer, for example, the question: ``What are the connections between Levallois-Perret and Africa?'' One possible answer comes by noticing that P. Balkany was the mayor of Levallois-Perret (as stated in the JSON document), and he owned the ``Dar Gyucy'' villa in Marrakesh (as shown in the relational table), which is in Morocco, which is in Africa (stated by the DBPedia subset). Another interesting connection in this graph is that Levallois-Perret appears in the same sentence
as the Centrafrican Republic in the Lib\'{e}ration snippet at the bottom right, which (as stated in DBPedia) is in Africa.

\mysection{Approach and outline}
\label{sec:outline}

We describe here the main principles of our approach, guided by the
requirements and constraints stated above. 

From requirement R1 (integral source preservation) it follows that
{\em all the structure and content of each dataset  is preserved}
in the integration graph, thus every detail of any dataset is mapped
to some of its nodes and edges. This requirement also leads us to 
{\em preserve the provenance} of each dataset, as well as {\em the
  links that may exist within and across datasets} before loading
them (e.g. interconnected HTML pages, JSON tweets
replying to one another, or RDF graphs referring to a shared resource). 
We term {\em primary nodes} the nodes created in our graph strictly
 based on the input dataset and their provenance; we detail their
 creation in Section~\ref{sec:primary}. 

From requirement R2 (ideally no user input), it follows that  {\em we
  must identify the opportunities to automatically link (interconnect)
  nodes}, even when they were not interconnected in their original
dataset, and even when they come from different datasets. We
achieve this at several levels:

First, we leverage and extend information extraction techniques
  to  {\em extract (identify) entities occurring in the labels of every node in every
    input  dataset}. For instance, ``Levallois-Perret'' is identified
  as a Location in the two datasets at right in
  Figure~\ref{fig:motiv-example} (JSON and Text). Similarly,  ``P. Balkany'', ``Balkany'', ``I. Balkany''
  occurring in the relational, JSON and Text datasets are extracted
  as Person entities. 
 Our method of entity extraction, in particular for the French
 language, is described in Section~\ref{sec:entity-extraction}.

Second, we {\em compare (match)} occurrences of entities extracted from the datasets, in order to determine when they refer to the same entity and thus should be interconnected. 
($i$)~Some entity occurrences we encounter refer to entities such as
Marrakech, Giverny etc. known in a {\em trusted
    Knowledge Base} (or KB, in short), such as DBPedia. Journalists
  may trust a KB for such general, non-disputed entities. We  {\em
    disambiguate} each entity occurrence, i.e., try to find the URI
  (identifier) assigned in  the KB to the entity referred to in this occurrence, and we
  {\em connect} the occurrence to the entity. Disambiguation enables, for instance, to
  connect an occurrence of ``Hollande'' to the country, 
  and another to the former
    French president. A common entity found in two datasets
    interconnects them. 
We describe the module we built for entity disambiguation
  for the French language (language constraint C3), based on AIDA~\cite{hoffart2011robust}, in
  Section~\ref{sec:disambig}. It is of independent interest, as it can
  be used outside of our context. 
($ii$)~On little-known entities (constraint
  C1), disambiguation fails (no URI is found); this is the case, e.g., of ``Moulin
  Cossy''. Combined with constraint C2
  (strict control on ingested sources) it leads to the lack of reliable IDs for many entities mentioned in the datasets. We strive to connect them, as soon as the several identical or at least {\em
    strongly similar} occurrences are found in the same or different datasets. We describe our approach for {\em comparing (matching)} occurrences in
  order to identify identical or similar pairs in
  Section~\ref{sec:matching}. Section~\ref{sec:storage} describes our
  persistent graph storage. 

For what concerns the keyword search problem, we formalize it in
Section~\ref{sec:pb-statement}, showing in particular that
requirement R4 leads to an explosion in the size of the search space,
compared to those studied in the literature. Section~\ref{sec:score}
discusses some favorable cost function properties exploited in the
literature, and why they do not apply to our case (constraint
C5). Based on this, Section~\ref{sec:algo} introduces our query
algorithm. 

We present experiments evaluating the quality and performance of our
algorithms in Section~\ref{sec:evaluation}. Our work pertains to several  areas, most notably data integration,  knowledge base
construction, and keyword search; we detail our positioning in
Section~\ref{sec:related}. 

\mysection{Primary graph construction from heterogeneous datasets}
\label{sec:primary}


We consider the following \emph{data models}: relational (including SQL databases, CSV files  etc.),  
RDF, 
JSON, XML, or  
HTML, 
and text. 
A dataset $DS=(db, prov)$ in our context is a pair, whose first component is a concrete data object: a relational database, or an RDF graph, or a  JSON, HTML, XML document, or a  CSV, text, or PDF file.  The second (optional) component $prov$ is the dataset provenance; we consider here that it is the URI from which the dataset was obtained, but this could easily be generalized. 

Let $A$ be an alphabet of words.
We define an \textbf{integrated graph} $G=(N,E)$ where $N$ is the set of nodes and $E$ the set of edges. We have $E\subseteq{N\times{N}\times{A^*}\times{[0,1]}}$, where $A^*$ denotes the set of (possibly empty) sequences of words, and the value in $[0,1]$ is the \textit{confidence}, reflecting the probability that the relationship between two nodes holds. 
Each node $n\in N$ has a label $\lambda (n)\in A^*$ and similarly each edge $e$ has $\lambda(e) \in A^*$. We use $\epsilon$ to denote the empty label. We assign to each node and edge a \textbf{unique ID}, as well as a \textbf{type}. We introduce the supported node types as needed, and write them in bold font (e.g., \textbf{dataset node}, \textbf{URI node}) when they are first mentioned; node types are important as they determine the quality and performance of matching (see Section~\ref{sec:matching}). Finally, we create unique dataset IDs and associate to each node its dataset's ID. 

Let $DS_i=(db_i, prov_i)$ be a dataset of any of the above models. The following two steps are taken regardless of $db_i$'s data model: 
First, we introduce a  \textbf{dataset node} $n_{DS_i}\in N$, which models the dataset itself (not its content). 
Second, if $prov_i$ is not null, we create an \textbf{URI node} $n_{prov_i}\in N$,  whose value is the provenance URI $prov_i$, and an edge $n_{DS_i}\xrightarrow{\text{cl:prov}}n_{prov_i}$, where cl:prov is a special edge label denoting provenance (we do not show these edges in the Figure to avoid clutter).

\begin{figure}[t!]
\centering
\includegraphics[width=.9\textwidth]{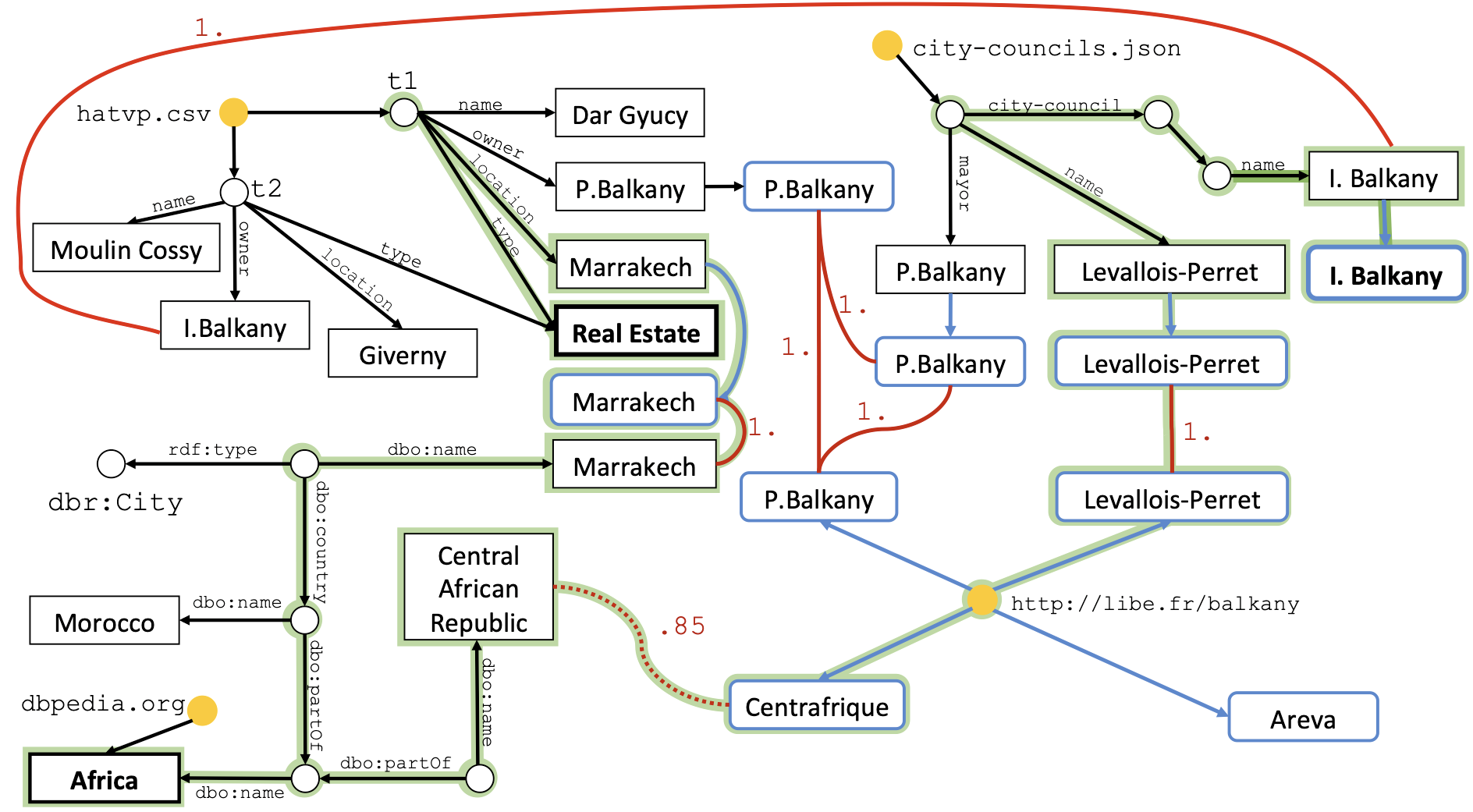}
\vspace{-3mm}
\caption{Integrated graph corresponding to the datasets of Figure~\ref{fig:motiv-example}.  An answer to the keyword query \{``I. Balkany'', Africa, Estate\} is highlighted in light green; the labels three keyword matches in this answer are in bold font.\label{fig:example-graph}}
\vspace{-7mm}
\end{figure}

Next, Section~\ref{sec:datasets-to-graph} explains how each type of dataset yields nodes and edges in $G$.  For illustration, Figure~\ref{fig:example-graph} shows the integrated graph resulting from the datasets in Figure~\ref{fig:motiv-example}. In Section~\ref{sec:graph-refine-optim}, we describe a set of techniques that improve the informativeness and the connectedness and decrease the size of $G$. 
Section~\ref{sec:complexity} discusses the complexity of the graph construction. 

\mysubsection{Mapping each dataset to the  graph}
\label{sec:datasets-to-graph}

All the edges whose creation is described in this section reflect the {\em structure} of the input datasets. Thus, their confidence is always $1.0$.

\noindent\textbf{Relational.}
\label{sec:vg-relational}
Let $db=R(a_{1},\ldots,a_{m})$ be a relation (table) (residing within an RDBMS, or ingested from a CSV file etc.)
A \textbf{table node} $n_R$  is created to represent $R$ (yellow node with label {\sf hatvp.csv} on top left in Figure~\ref{fig:example-graph}).
Let $t\in R$ be a tuple of the form  $ (v_{1},\ldots,v_{m})$ in $R$. 
A \textbf{tuple node} $n_t$ is created for $t$, with an empty label, e.g., $t_1$ and $t_2$ in Figure~\ref{fig:example-graph}. 
For each non-null attribute $v_i$ in $t$, a \textbf{value node} $n_{v_i}$ is created, together with an edge from $n_t$ to $n_{v_{i}}$, 
labeled $a_{i}$ 
 (for example,  the edge labeled {\sf owner} at the top left in Figure~\ref{fig:example-graph}). To keep the graph readable, confidence values of 1 are not shown.
Moreover, for any two relations $R,R'$ for which we know that attribute $a$ in $R$ is a foreign key referencing $b$ in $R'$, and for any tuples 
 $t \in R$, $t' \in R'$ such that $t.a = t'.b$, the graph comprises an edge from $n_{t}$ to $n_{t'}$ with confidence $1$. This graph modeling of relational data has been used for keyword search in relational databases~\cite{DBLP:conf/vldb/HristidisP02}.

\noindent\textbf{RDF.}
\label{sec:vg-rdf}
The mapping from an RDF graph to our graph is the most natural. 
Each node in the RDF graph becomes, respectively, a URI node or a value node in $G$, and each 
RDF triple becomes an edge in $E$. At the bottom left in Figure~\ref{fig:example-graph} appear some edges resulting from our DBPedia snippet.

\noindent\textbf{Text.} We model a text document very simply, as a node having a sequence of children, where each child is a segment of the text (e.g., the four nodes connected by the libe.fr node at the bottom right of Figure~\ref{fig:example-graph}). Currently, each segment is a phrase; we found this a convenient granularity for users inspecting the graph. Any other segmentation could be used instead. 

\noindent\textbf{JSON.}
\label{sec:vg-json}
As customary, we view a JSON document as a tree, with nodes that are either \textbf{map nodes}, \textbf{array nodes} or value nodes. We map each node into a node of our graph and create an edge for each parent-child relation. Map and array nodes have the empty label $\epsilon$. Attribute names within a map become edge labels in our graph. Figure~\ref{fig:example-graph} at the top right shows how a JSON document's nodes and edges are ingested in our graph. 


\noindent\textbf{XML.}
The ingestion of an XML document is very similar to that of JSON
ones. XML nodes are either \textbf{element}, or \textbf{attribute}, or
values nodes. As customary when modeling XML, value nodes are either text children of elements or values of their attributes. 

\noindent\textbf{HTML.} 
\label{sec:vg-html}
An HTML document is treated very similarly to an XML one. In particular, when an HTML document contains a hyperlink of the form \textsf{\small <a href=''http://a.org''>...</a>}, we create a node labeled ``a''  and another labeled ``http://a.org'', and connect them through an edge labeled ``href''; this is the common treatment of element and attribute nodes. However, we detect that a child node satisfies a URI syntax, and {\em recognize (convert) it into a URI node}. This enables us to preserve links across HTML documents ingested together in the same graph,  with the help of node comparisons (see Section~\ref{sec:matching}).

\noindent\textbf{PDF.}  A trove of useful data is found in the PDF format. To take advantage of their content, we have developed a PDF scrapper which transforms a PDF file into: ($i$)~a JSON file, typically containing all the text found in the PDF document; ($ii$)~if the PDF file contains bidimensional (2d) tables, each table is extracted in a separate RDF file, following an approach introduced in~\cite{cao:hal-01583975}, which preserves the logical connections of each data cell with its closest. 
The JSON and possibly the RDFs thus obtained are ingested as explained above; moreover, from the dataset node of each such dataset $d_i$, we add an edge labeled cl:extractedFromPDF, whose value is the URI of the PDF file. Thus, the PDF-derived datasets are all interconnected through that URI. 

\mysubsection{Refinements and optimizations}
\label{sec:graph-refine-optim}

\noindent\textbf{Value node typing.} The need to recognize particular types of values goes beyond identifying URIs in HTML.  URIs also frequently occur in JSON (e.g., tweets), in  CSV datasets etc. 
Thus, we examine each value to see if it follows the syntax of a URI and, if so, convert the value node into a URI one, regardless of the nature of the dataset from which it comes.
More generally, other categories of values can be recognized in order to make our graphs more meaningful. Currently, we similarly recognize \textbf{numeric nodes}, \textbf{date nodes}, \textbf{email address nodes} and \textbf{hashtag nodes}. 


\noindent\textbf{Node factorization.} The graph resulting from the ingestion of a JSON, XML, or HTML document, or one relational table, is a tree; any value (leaf) node is reachable by a finite set of label paths from the dataset node. For instance, in an XML document,  two value nodes labeled ``Paris'' may be  reachable on the paths employee.address.city, while another is on the path headquartersCity. Graph creation as described in Section~\ref{sec:datasets-to-graph} creates three value nodes labeled ``Paris''; we call this \textbf{per-occurrence value node creation}. Instead,  \textbf{per-path} creation leads to a single node for all occurrences of ``Paris'' on the paths employee.address.city and employee.headquartersCity.
We have also experimented with \textbf{per-dataset} value node creation, which in the above example  creates a single ``Paris'' node, and with \textbf{per-graph}, where a single ``Paris'' value node is created in a graph, regardless of how many times ``Paris'' appears across all the datasets. Per-graph is consistent with the RDF data model, where each literal denotes one node. 

Factorization turns a tree-structured dataset into a directed acyclic graph (DAG); it reduces the number of nodes, and increases the graph connectivity. 
Factorization may introduce {\em erroneous connections}. For instance, constants such as {\em true} and {\em false} appear in many contexts, yet this should not lead to connecting all nodes having an attribute whose value is {\em true}. Another example are named entities, which should be first disambiguated.

To prevent such erroneous connections, we have heuristically identified a set of  \textbf{values which should  not be factorized} even with {\em per-path}, {\em per-dataset} or {\em per-graph} value node creation. Beyond {\em true} and {\em false} and {\em named entities}, this currently includes {\em integer numeric node labels written on less than $4$ digits}, the rationale being that small integers tend to be used for ordinals, while numbers on many digits could denote years, product codes, or other forms of identifiers. This simple heuristic could be refined. 
\textbf{Null codes}, or strings used to signal missing values, e.g., ``N/A'', ``Unknown'', should not be factorized, either. As is well-known from database theory, nulls should not lead to joins (or connections, in our case). 
Nodes with such labels will never lead to connections in the graph: they are not factorized, and they are not compared for similarity (see Section~\ref{sec:matching}). This is why we currently require user input on the null codes, and assist them by showing the most frequent constants, as null codes, when they occur, tend to be more frequent than real values (as our experiments illustrate in Section~\ref{sec:exp-construction}).  Devising an automated approach toward detecting null codes in the data is an interesting avenue for future work.

\mysubsection{Complexity of the graph construction}
\label{sec:complexity}

 Summing up the above  processing stages, the worst case complexity of
ingesting a set of datasets in a ConnectionLens graph $G=(N,E)$ is of the
form:

\vspace{-1.5mm}
\begin{center}
$c_1\cdot |E| + c_2\cdot |N| + c_3 \cdot |N_e| +  c_4 \cdot |N|^2$
\end{center}
\vspace{-1.5mm}

In the above, the constant factors are explicitly present (i.e., not
wrapped in an $O(\ldots)$ notation) as the differences between them
are high enough to significantly impact the overall construction time (see Section~\ref{sec:exp-ner} and
\ref{sec:exp-disambiguation}). Specifically: 
 $c_1$ reflects the (negligible) cost of creating each edge using the
 respective data parser, and the (much higher) cost of storing it;
$c_2$  reflects the cost to store a node in the database, and to
invoke the entity extractor on its label, if it is not $\epsilon$;
$N_e$ is the number of entity nodes found in the graph, and $c_3$ is
the cost to disambiguate each entity; finally, the last component
reflects the worst-case complexity of node matching, which may be
quadratic in the number of nodes compared. The constant $c_4$ reflects
the cost of recording on disk that the two nodes are equivalent (one
query and one update) or similar (one update). 

Observe that the number of value nodes (thus
$N$) is impacted by the node creation policy; $N_e$ (and, as we show
below, $c_3$) depend on the entity extractor module used.

\mysection{Named-Entity Recognition}
\label{sec:entity-extraction}

We enrich our graphs by leveraging Machine Learning (ML) tools for Information Extraction.

{\em Named entities} (NEs) \cite{nadeau2007ner} are words or phrases which, together, designate certain real-world entities. Named entities include common concepts such as people, organizations, and locations. 
The {\em Named-Entity Recognition} (NER) task consists of $(i)$~identifying  NEs in a natural language text, and  ($ii$)~classifying them according to a pre-defined set of NE types. 
Let $n_t$ be a text node. We feed $n_t$ as input to a NER module and create, for each entity occurrence $E$ in $n_t$, an \textbf{entity occurrence node} (or entity node, in short) $n_E$; as explained below, we extract \textbf{Person, Organization} and \textbf{Location} entity nodes. Further, we add an edge from $n_t$ to $n_E$ whose label is cl:extract$T$, where $T$ is the type of $E$, and whose confidence is $c$, the {\em confidence of the extraction}.  In Figure~\ref{fig:example-graph}, the blue, round-corner rectangles {\sf Centrafrique, Areva, P. Balkany, Levallois-Perret} correspond to the entities recognized from the  text document, while the {\sf Marrakech} entity is extracted from the identical-label value node originating from the CSV file. 

\vspace{3mm}
\noindent\textbf{Named-Entity Recognition} 
We describe here the NER approach we devised for our framework, for English and French.
While we have used Stanford NER~\cite{finkel2005incorporating} in~\cite{Chanial2018}, we have subsequently developed a more performant module based on the Deep Learning Flair NLP framework~\cite{akbik2019flair}. 
Flair and similar frameworks rely on {\em embedding} words into vectors in a multi-dimensional space. Traditional word embeddings, e.g., Word2Vec \cite{mikolov2013efficient}, Glove \cite{pennington2014glove} and fastText \cite{bojanowski2017enriching}, are \emph{static}, meaning that a word's representation does not depend on the context where it occurs. New embedding techniques are \emph{dynamic}, in the sense that the word's representation also depends on its context. In particular, the Flair dynamic embeddings~\cite{akbik2018contextual} achieve state-of-the-art NER performance. The latest Flair architecture~\cite{akbik2019flair} facilitates {\em combining} different types of word embeddings, as a better performance might be achieved by combining dynamic with static word embeddings. 

For English, we rely on a  model\footnote{https://github.com/flairNLP/flair} pre-trained using the English CoNLL-2003\footnote{https://www.clips.uantwerpen.be/conll2003/ner/} news articles dataset. 
The model combines Glove embeddings~\cite{pennington2014glove} and so-called {\em forward and backward pooled} Flair embeddings, that evolve across subsequent extractions. 
As such a model was missing for French, we trained a Flair one on WikiNER~\cite{nothman2013learning},  a multilingual NER dataset automatically created using the text and structure of Wikipedia.
The dataset contains $132$K sentences, $3.4$M tokens and $216$K named-entities, including $74$K Person, $116$K Location and $25$K Organization entities. 
The model uses stacked forward and backward French Flair embeddings with French fastText~\cite{bojanowski2017enriching} embeddings.

\noindent\textbf{Entity node creation.} Similarly to the discussion about value node factorization (Section~\ref{sec:graph-refine-optim}), we have the choice of creating an entity node $n_E$ of type $t$ once per occurrence, or (in hierarchical datasets) {\em per-path}, {\em per-dataset} or {\em per-graph}.   
We adopt the per graph method, with the mention that we will create one entity node for each disambiguated entity and one entity node for each non-disambiguated entity.

\mysection{Entity disambiguation}
\label{sec:disambig}

Some (but not all) entity nodes extracted from a dataset as an entity of type $T$ may correspond to an entity (resource) described in a trusted knowledge base (KB) such as DBPedia or Yago. When this is the case, this allows: ($i$)~resolving {\em ambiguity} to make a more confident decision about the entity, e.g., whether the entity node ``Hollande'' refers to the former president or to the country; ($ii$)~tackling {\em name variations}, e.g., two Organization entities labeled ``Paris Saint-Germain Football Club'' and ``PSG'' are linked to the same KB identifier, and ($iii$)~if this is desired, {\em enriching} the dataset with a certain number of facts the KB provides about the entity. 

{\em Named entity disambiguation} (NED, in short, also known as entity linking) is the process of assigning a unique identifier (typically, a URI from a KB) to each named-entity present in a text.  We built our NED module based on  AIDA~\cite{hoffart2011robust}, part of the Ambiverse\footnote{\url{https://www.mpi-inf.mpg.de/departments/databases-and-information-systems/research/ambiverse-nlu/}} framework; AIDA maps entities to resources in YAGO 3~\cite{DBLP:conf/cidr/MahdisoltaniBS15} and Wikidata~\cite{wikidata}.
Our work consisted of ($i$)~adding  support for French (not present in Ambiverse), and ($ii$)~integrating our own NER module (Section~\ref{sec:entity-extraction}) within the Ambiverse framework.


For the first task, in collaboration with the maintainers of Ambiverse\footnote{\url{https://github.com/ambiverse-nlu/ambiverse-nlu\#maintainers-and-contributors}}, we built a new dataset for French, containing the information required for AIDA. The dataset consists of entity URIs, information about entity popularity (derived from the frequency of entity names in link anchor texts within Wikipedia),  and entity context (a set of weighted words or phrases that co-occur with the entity), among others. This information is language-dependent and was computed from the French Wikipedia.

For what concerns the second task,  Ambiverse takes an input text and passes it through a text processing pipeline consisting of \emph{tokenization} (separating words), \emph{part-of-speech (POS) tagging}, which identifies nouns, verbs, etc.,  NER, and finally NED. Text and annotations are stored and processed in Ambiverse using the UIMA standard\footnote{\url{https://uima.apache.org/doc-uima-why.html}}.  A central UIMA concept is the {\em Common Annotation Scheme} (or CAS); in short, it encapsulates the document analyzed, together with all the annotations concerning it, e.g., token offsets, tokens types, etc.  In each Ambiverse module, the CAS object containing the document receives new annotations, which are used by the next module. For example, in the tokenization module, the CAS initially contains only the original text; after tokenization, the CAS also contains token offsets. 

To integrate our Flair-based extractor (Section~\ref{sec:entity-extraction}), we deployed a \textbf{new Ambiverse processing pipeline}, to which we pass as input {\em both the input text and the extracted entities}. 


\mysection{Node matching}
\label{sec:matching}



This section presents our fourth and last method for \textbf{identifying and materializing connections} among nodes from the same or different datasets of the graph. 
Recall that ($i$)~{\em value nodes with identical labels can be fused (factorized)} (Section~\ref{sec:graph-refine-optim});
($ii$)~nodes become connected as {\em parents of a common extracted entity} (Section~\ref{sec:entity-extraction}); 
($iii$)~entity nodes with different labels can be interconnected {\em through a common reference entity} when NED returns the same KB entity (Section~\ref{sec:disambig}).  Other pairs of nodes with similar labels, that may still need to be connected, are: {\em entity pairs} where disambiguation, which is context-dependent, returned no result for one or for both {\em (entity, value)} pairs where the value corresponds to a node from which no entity was extracted (the extraction is context-dependent and may also have some misses); and {\em value pairs}, such as numbers, dates, and (if desired) identical texts. Comparing texts is useful, e.g., to compare social media posts when their topics (short strings) and/or body (long string) are very similar. 

When a comparison finds two nodes with \emph{very similar labels}, we create an edge labeled \textbf{cl:sameAs}, whose confidence is the similarity between the two labels. In Figure~\ref{fig:example-graph}, a dotted red edge (part of the subtree highlighted in green) with confidence
$.85$ connects the ``Central African Republic'' RDF literal node with the
``Centrafrique'' Location entity extracted from the text. 

When a comparison finds two nodes with \emph{identical labels}, one could unify them, but this raises some modeling issues, e.g., when a value node from a dataset $d_1$ is unified with an entity encountered in another dataset  $d_2$. Instead, we {\em conceptually} connect the nodes with sameAs edges whose confidence is $1.0$. These edges are drawn in solid red lines in Figure~\ref{fig:example-graph}. Nodes  connected by a $1.0$ sameAs edge are also termed {\em equivalent}. 
Note that conceptual equivalence edges are not stored; instead, the information about $k$ equivalent nodes is stored using $O(k)$ space, as we explain in Section~\ref{sec:storage}.

Inspired by the data cleaning literature, we start by {\em normalizing} node labels, e.g., person names are analyzed to identify first names, last names, civility prefixes such as ``Mr.'', ``Dr.'', and a normalized label is computed as ``Firstname Lastname''. Subsequently, our approach for matching is {\em set-at-a-time}. More precisely, we form node group pairs $(Gr_1, Gr_2)$, and we compare pair of labels  using the {\em similarity function} known to give the best results (in terms of matching quality) for those groups.  The Jaro measure \cite{jaro1989sim} gives good results for {\em short strings}~\cite{doan2012int} and is applied to compute the similarity between pairs of entities of the same type recognized by the entity extractor (i.e., person names, locations and organization names which are typically described by short strings). 

\mysection{Graph storage}
\label{sec:storage}
Currently, our graphs are stored in a
relational database, which is accessed through JDBC. This solution has the advantage of relying on
a standard (JDBC) backed by mature, efficient and free tools,
such as PostgreSQL,
which runs on a variety of platforms (requirement R3 from Section~\ref{sec:intro}).

The table  \textbf{Nodes(\underline{id}, label, type, datasource, label,
normaLabel, representative)} stores the basic attributes of a node,
the ID of its data source, its normalized label, and the ID of its
representative. For nodes not equivalent to any other, the
representative is the ID of the node itself.  As explained previously, the
{\em representative} attribute allows encoding information about
equivalent nodes.
 Table \textbf{Edges(\underline{id}, source, target, label, datasource, confidence)} stores the
 graph edges derived from the data sources, as well as the extraction
 edges connecting entity nodes with the parent(s) from which they have
 been extracted. Finally, the \textbf{Similar(source, target,
 similarity)} table stores a tuple for each similarity comparison
 whose result is above the threshold but  less than $1.0$.
The pairs of nodes to be compared for similarity are retrieved by
means of SQL queries. 


This relational store is encapsulated behind a Graph interface, which
can be easily implemented differently to take advantage of other
storage engines. When creating a graph, storage is far from being the
bottlenck (as our experiments in Section~\ref{sec:exp-construction} show); on
the contrary, when querying the graph (see below), the relational store incurs a relatively high access cost
for each edge. We currently mitigate this
problem using a memory cache for nodes and edges, within the ConnectionLens application.

\section{Querying the graph}
\label{sec:pb-statement}

We formalize the keyword search problem over a  graph built out of heterogeneous datasets as previously described.

\subsection{Search problem}
\label{sec:search-problem}

We consider a graph $G=(N, E)$ and we denote by $\mathcal{L}$ the set of all the labels of $G$ nodes, plus the empty label $\epsilon$ (see Figure~\ref{fig:example-graph}). Let $W$ be the set of {\em keywords}, obtained by stemming the label set $\mathcal{L}$; a {\em search query} is a set of keywords $Q=\{w_1,...,w_m\}$, where $w_i\in W$.
We define an \textbf{answer tree} (AT, in short) as a set $t$ of  $G$ edges which ($i$)~together, form a tree (each node is reachable from any other through exactly one path), ($ii$)~for each $w_i$, contain at least one node whose label matches $w_i$. Here, the edges are \textbf{considered undirected}, that is: $n_1\xrightarrow{a}n_2\xleftarrow{b}n_3\xrightarrow{c}n_4$ is a sample AT,  such that for all $w_i \in Q$, there is a node  $n_i\in t$ such that $w_i \in \lambda(n_i)$.




We treat the edges of $G$ as undirected when defining the AT in order to allow more query results, on a graph built out of heterogeneous content whose structure is not well-known to users.
Further, we are interested in  \textbf{minimal} answer trees, that is:
(i) removing an edge from the tree should make it lack one or more of the query keywords $w_i$;
(ii) if a query keyword $w_i$ matches the label of more than one node in the answer tree, then all these matching nodes must be equivalent.

Condition~(ii) is specific to the graph we consider, built from {\em several data sources connected by equivalence or similarity edges}. In classical graph keyword search problems, each query keyword is matched {\em exactly once} in an answer (otherwise, the tree is considered non-minimal). In contrast, our answer trees \emph{may need to traverse equivalence edges}, and if $w_i$ is matched by one node connected by such an edge, it is also matched by the other. For instance, consider the three-keyword query ``Gyucy Balkany Levallois'' in Figure~\ref{fig:example-graph}:  the keyword Balkany is matched by the two nodes labeled ``P. Balkany'' which are part of the answer.

As a counter-example, consider the query ``Balkany Centrafrique'' in Figure~\ref{fig:example-graph}, assuming the keyword Centrafrique is also matched in the label ``Central African Republic''\footnote{This may be the case using a more advanced indexing system that includes some natural language understanding, term dictionaries etc.}.
Consider the tree that connects a ``P. Balkany'' node with ``Centrafrique'', and also traverses the edge between ``Centrafrique'' and ``Central African Republic'': this tree is not minimal, thus it is not an answer. The intuition for rejecting it is that  ``Centrafrique'' and ``Central African Republic'' are not necessarily equivalent (we have a similarity, not an equivalence edge), therefore the query keyword ``Centrafrique'' is matched by two potentially different things in this answer, making it hard to interpret.


A direct consequence of minimality is that {\em in an answer, each and every leaf matches a query keyword}.
A graph may hold several minimal answer trees for a given query. We consider available a {\em scoring function} which assigns a higher value to more interesting answer trees (see Section~\ref{sec:score}).

\textbf{Problem Statement.}
Given the graph $G$ built out of the datasets  $\mathcal{D}$ and a query $Q$,  return the $k$ highest-score minimal answer trees.~\qed

An AT may potentially span over the whole graph, (also) because it can traverse $G$ edges in any direction;  this makes the problem challenging.




\subsection{Search space and complexity}
\label{sec:steiner}

\newcommand\pbma{$\diamond$}
\newcommand\pbmb{$\rhd$}
\newcommand\pbmc{$\lhd$}
\newcommand\pbmd{$\circ$}
\newcommand\pbme{$\Box$}

The problem that we study is related to the (Group) Steiner Tree Problem, which we recall below.

Given a graph $G$ with weights (costs) on edges, and a set of $m$ nodes $n_1,\ldots,n_m$, the \emph{Steiner Tree Problem (STP)} \cite{garey2011} consists of finding the smallest-cost tree in $G$ that connects all the nodes together. We could answer our queries by solving one STP problem for each combination of nodes matching the keywords $w_1,\ldots,w_m$. However, there are several obstacles left: (\pbma)~STP is a known NP-hard problem in the size of $G$, denoted $|G|$; (\pbmb)~as we consider that each edge can be taken in the direct or reverse direction, this amounts to ``doubling'' every edge in $G$. Thus, our search space is  \textbf{$2^{|G|}$ larger than the one of the STP, or that considered in similar works}, discussed in Section~\ref{sec:related}. This is daunting even for small graphs of a few hundred edges;   (\pbmc)~we need the $k$ smallest-cost trees, not just one; (\pbmd)~each keyword may match several nodes, not just one.

The closely related {\em Group STP} (GSTP, in short) \cite{garey2011} is: given $m$ {\em sets of nodes} from $G$, find the minimum-cost subtree connecting one node from each of these sets. GSTP does not raise the problem (\pbmd), but still has all the others.

In conclusion, the complexity of the problem we consider is extremely high. Therefore, solving it fully is unfeasible for large and/or high-connectivity graphs. Instead, our approach is:
($i$)~{\em Attempt to find all answers from the smallest} (fewest edges) {\em to the largest}. 
 Enumerating small trees first is both a practical decision (we use them to build larger ones) and fits the intuition that we should not miss small answers that a human could have found manually.
However, as we will explain, we still ``opportunistically'' build some trees before exhausting the enumeration of smaller ones, whenever this is likely to lead faster to answers. The strategy for choosing to move towards bigger instead of smaller tress leaves rooms for optimizations on the search order.
($ii$)~{\em Stop at a given time-out or when $m$ answers have been found}, for some $m\geq k$;
($iii$)~{\em Return the $k$ top-scoring answers} found.


\section{Scoring answer trees}
\label{sec:score}

We now discuss how to evaluate the quality of an
answer. Section~\ref{sec:generic-score} introduces the general notion
of score on which we base our approach. Section~\ref{sec:specif}
describes the metric that we attach to edges in order to
instantiate this score, and Section~\ref{sec:concrete-score}
details the actual score function we used.

\subsection{Generic score function}
\label{sec:generic-score}
We have configured our problem setting to allow {\em any scoring function}, which enables the use of different scoring schemes fitting the requirements of different users.
As a consequence, this approach allows us to study the interaction of the scoring function with different properties of the graph.

Given an answer tree $t$ to a query $Q$, we consider a score function consisting of (at least) the following two components.
First, the {\em matching score} $ms(t)$, which reflects the quality of the answer tree, that is, how well its leaves match the query terms.
Second, the {\em connection score} $cs(t)$, which reflects the quality of
  the tree connecting the edges.
  Any formula can be used here, considering the number of edges, the confidence or any other property attached to edges, or a query-independent property of the nodes, such as their PageRank or betweenness centrality score etc.

The score of $t$ for $Q$, denoted $s(t)$, is computed as a
combination of the two independent components $ms(t)$ and $cs(t)$.
Popular combinations functions (a weighted sum, or  product etc.) are monotonous in both components, however, our framework does not require monotonicity. Finally, both $ms(t)$ and $cs(t)$ can be tuned based on a given user's preferences, to personalize the score, or make them evolve in time through user feedback etc.

\mysubsection{Edge specificity}\label{sec:specif}
We  now describe a metric on edges, which we used (through the
connection score $cs(t)$) to favor edges that are ``rare'' for both nodes they
connect. This metric was inspired by our experiments with real-world
data sources, and it helped return interesting answer trees in our
experience. 


For a given node $n$ and label $l$, let $\edgesin{n}{l}$ be the number of $l$-labeled edges entering $n$, and $\edgesout{n}{l}$ the number of $l$-labeled edges exiting $n$.

\noindent The \textbf{specificity} of an edge $e=n_1\xrightarrow{l}n_2$ is defined as:

\begin{center}
\va\va
$s(e)=2/(\edgesout{n_1}{l} + \edgesin{n_2}{l})$.
\va\va
\end{center}

Specificity is $1.0$ for edges that are ``unique'' for both their source and their target, and decreases when the edge does not ``stand out'' among the edges of these two nodes.
For instance,  the city council of Levallois-Perret comprises only one mayor (and one individual cannot be mayor of two cities in France).
Thus, the edge from the city council to P.~Balkany has a specificity of $2/(1.0+1.0)=1.0$. In contrast, there are 54 countries in Africa (we show only two), and each country is in exactly one continent;  thus,
the specificity of the {\sf dbo:partOf } edges in the DBPedia fragment, going from the node named Morocco (or the one named Central African Republic) to the node named Africa is $2/(1+54)\simeq .036$.

\subsection{Concrete score function}
\label{sec:concrete-score}

We have implemented the following prototype scoring function in our system.
For an answer $t$ to the query $Q$, we compute the matching score $ms(t)$  as the \emph{average}, over all query keywords $w_i$, of the similarity between the $t$ node matching $w_i$ and the keyword $w_i$ itself; we used the edit distance.

We compute the connection score $cs(t)$ based on edge confidence, on one hand, and edge specificity on the other. We {\em multiply} the confidence values, since we consider that uncertainty (confidence $<1$) multiplies; and we also {\em multiply}  the specificities of all edges in $t$, to discourage many low-specificity edges. Specifically, our score is computed as:



\begin{center}
$score(t,Q)=\alpha \cdot ms(t,Q) + \beta \cdot \prod_{e\in  E} c(e) + (1 - \alpha - \beta) \cdot \prod_{e\in E} s(e)$
\end{center}

\noindent where $\alpha$, $\beta$ are parameters of the system such that $0\leq \alpha, \beta <1$ and $\alpha +\beta\leq 1$.

\section{Answering keyword queries}
\label{sec:algo}

We now present our approach for computing query answers, on the graph which integrates the heterogeneous datasets.

\subsection{\grow\ and  \mergecur}
\label{sec:algorithm}

Our algorithm relies on concepts from prior literature~\cite{dpbf,blinks} while exploring many more trees.
Specifically, it starts from the sets of nodes $N_1,\ldots,N_m$ where the nodes in $N_i$ all match the query keyword $w_i$; each node $n_{i,j}\in N_i$ forms a one-node partial tree.
For instance, in Figure~\ref{fig:example-graph}, one-node trees are built from the nodes with boldface text, labeled ``Africa'', ``Real Estate'' and ``I. Balkany''.
We identify two transformations that can be applied to form increasingly larger trees, working toward query answers:

\noindent\textbf{\grow($t,e$)}, where $t$ is a tree, $e$ is an edge \emph{adjacent to the root} of $t$, and $e$ does not close a loop with a node in $t$, creates a new tree $t'$ having all the edges of $t$ plus $e$; the root of the new tree is the other end of the edge $e$. For instance, starting from the node labeled ``Africa'', a \grow\ can add the edge labeled {\small \texttt{dbo:name}}.

\noindent\textbf{\mergecur($t_1,t_2$)}, where $t_1,t_2$ are trees with the same root, whose other nodes are disjoint, and
matching disjoint sets of keywords, creates a tree $t''$ with the same root and with all edges from $t_1$ and $t_2$. 
Intuitively, \grow\ moves away from the keywords, to explore the graph; \mergecur\ fuses two trees into one that matches more keywords than both $t_1$ and $t_2$.

In a {\em single-dataset} context, \grow\ and \mergecur\  have the following properties.
($gm_1$)~\grow\ alone is \textbf{complete} (guaranteed to find all answers) for $k=1,2$ only; for higher $k$, \grow\ and \mergecur\, together are complete.
($gm_2$)~Using \mergecur\ steps helps to find answers faster than using just \grow~\cite{blinks}: partial trees, each starting from a leaf that matches a keyword, are merged into an answer as soon as they have reached the same root.
($gm_3$)~An answer can be found through \textbf{multiple combinations of \grow\ and \mergecur}. For instance, consider a linear graph $n_1\rightarrow n_2 \rightarrow \ldots n_p$ and the two-keyword query $\{a_1, a_p\}$ where $a_i$ matches the label of $n_i$. The answer is obviously the full graph. It can be found: starting from $n_1$ and applying $p-1$ \grow\ steps; starting from $n_p$ and applying $p-1$ \grow\ steps; and in $p-2$ ways of the form \mergecur(\grow(\grow\ldots), \grow(\grow\ldots)), each merging in an intermediary node $n_2,\ldots,n_{p-1}$. These are all the same according to our definition of an answer (Section~\ref{sec:search-problem}), which does not distinguish a root in an answer tree; this follows users' need to know how things are connected, and for which the tree root is irrelevant.

\subsection{Adapting to multi-datasets graphs}
The changes we brought for our harder problem (bidirectional edges and multiple interconnected datasets) are as follows.

\noindent\textbf{1. Bidirectional growth.} We allow \grow\ to traverse an edge both going from the source to the target, and going from the target to the source.

\noindent\textbf{2. Many-dataset answers.} As defined in a single-dataset scenario, \grow\ and \mergecur\ do not allow to connect multiple datasets. To make that possible, we need to enable one, another, or both to also traverse similarity and equivalence edges (shown in solid or dotted red lines in Figure~\ref{fig:example-graph}). We decide to simply extend \grow\ to allow it to traverse not just data edges, but also {\em similarity} edges between nodes of the same or different datasets.
We handle {\em equivalence} edges as follows: 

\newcommand{\gtr}{\textsc{Grow2Rep}}
\noindent\textbf{\grow-to-representative}
Let $t$ be a partial tree developed during the search, rooted in a node $n$, such that the  representative of $n$ is a node $n_{rep}\neq n$.  \gtr\ creates a new tree by adding to $t$ the edge $n\xrightarrow{\equiv}n_{rep}$; this new tree is rooted in $n_{rep}$. If $n$ is part of a group of $p$ equivalent nodes,  only  {\em one} \gtr\ step is possible from $t$, to the unique representative of $n$; \gtr\ does not apply again on \gtr($t$), because the root of this tree is $n_{rep}$, which is its own representative.

Together, \grow, \gtr\ and \mergecur\ enable finding answers that span multiple data sources, as follows:
\grow\ allows exploring data edges within a dataset, and similarity edges within or across datasets.
\gtr\ goes from a node to its representative when they differ; the representative may be in a different dataset.
\mergecur\ merges trees with a same root: when that root represents $p$ equivalent nodes, this allows connecting partial trees, including \gtr\ results, containing nodes from different datasets. Thus, \mergecur\ can build trees spanning multiple datasets.

One potential performance problem remains. Consider again $p$ equivalent nodes $n_1,\ldots,n_p$; assume without loss of generality that their representative is $n_1$. Assume that during the search, a tree $t_i$ is created rooted in each of these $p$ nodes. \gtr\ applies to all but the first of these trees, creating the trees $t_2', t_3', \ldots, t_p'$, all rooted in $n_1$. Now, \mergecur\ can merge any pair of them, and can then repeatedly apply to merge three, then four such trees etc., as they all have the same root $n_1$. The exponential explosion of \grow\ trees, avoided by introducing \gtr, is still present due to \mergecur.

We solve this problem as follows.
Observe that in an answer, a  {\em path of two or more equivalence edges} of the form $n_1\xrightarrow{\equiv}n_2\xrightarrow{\equiv}n_3$ such that {\em a node internal to the path}, e.g. $n_2$, {\em has no other adjacent edge}, even if allowed by our definition, is \emph{redundant}. Intuitively, such a node brings nothing to the answer, since its neighbors, e.g., $n_1$ and $n_3$, could have been connected directly by a single equivalence edge, thanks to the transitivity of equivalence. We call {\em non-redundant} an answer that does not feature any such path, and decide to \textbf{search for non-redundant answers} only.


The following properties hold on non-redundant answers:

\begin{property}
There exists a graph $G$ and a $k$-keyword query $Q$ such that a non-redundant answer contains $k-1$ adjacent equivalence edges
(edges that, together, form a single connected subtree).
\end{property}

\begin{wrapfigure}{L}{0.6\textwidth}
\vspace{-5mm}
\centering
\includegraphics[width=.58\textwidth]{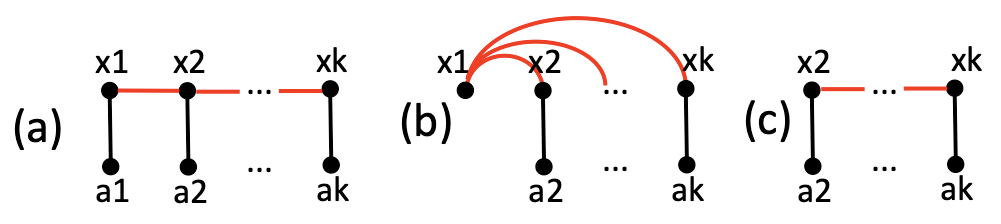}
\vspace{-2mm}
\caption{Sample graph and answer trees.\label{fig:prop-example}}
\vspace{-5mm}
\end{wrapfigure}

We prove this by exhibiting such an instance.  Let $G$ be a graph of $2k$ nodes shown in Figure~\ref{fig:prop-example} (a),  such that all the $x_i$ are equivalent, 
and consider the $k$-keyword query $Q=\{a_1,\ldots,a_k\}$ (each keyword matches exactly the respective $a_i$ node).  An answer needs to traverse all the $k$ edges from $a_i$ to $x_i$, and then connect the nodes $x_i,\ldots,x_k$; we need $k-1$ equivalence edges for this.

Next, we show:

\begin{property}\label{prop:2}
Let $t$ be a non-redundant answer to a query $Q$ of $k$ keywords. A group of adjacent equivalence edges contained in $t$  has at most $k-1$ edges.
\end{property}

We prove this by induction over $k$.
For $k=1$, each answer has $1$ node and $0$ edge (trivial case).

Now, consider this true for $k$ and let us prove it for $k+1$. Assume by contradiction that a non-redundant answer $t_Q$ to a query $Q$ of $k+1$ keywords comprises $k+1$ adjacent equivalence edges. Let $Q'$ be the query having only the first $k$ keywords of $Q$, and $t'$ be a subtree of $t$ that is a non-redundant answer to $Q'$:
\begin{itemize}
\item $t'$ exists, because $t$ connects all $Q$ keywords, thus also the $Q'$ keywords;
\item $t'$ is non-redundant, because all its edges are in the (non-redundant) $t$.
\end{itemize}

By the induction hypothesis, $t'$ has at most  $k-1$ adjacent equivalence edges. This means that there are {\em two adjacent equivalent edges} in $t\setminus t'$.  

\begin{enumerate}
\item If these edges, together, lead to two distinct leaves of $t$, then $t$ has {\em two} leaves  not in $t'$. This is not possible, because by definition of an answer, $t$ has $k+1$ leaves (each matching a keyword) and similarly $t'$ has $k$ leaves.
\item It follows, then, that the two edges lead to a single leaf of $t$, therefore the edges form a redundant path. This contradicts the non-redundancy of $t$, and
concludes our proof.
\end{enumerate}

Property~\ref{prop:2} gives us an important way to control the exponential development of trees due to $p$ equivalent nodes. \grow, \gtr\ and \mergecur, together, can generate trees with up to $k$ (instead of $k-1$) adjacent equivalence edges. 
This happens because \gtr\ may ``force'' the search to visit the representative of a set of $k$ equivalent nodes (see Figure~\ref{fig:prop-example}(b), assuming $x_1$ is the representative of all the equivalent $x_i$s, and the query $\{a_2,\ldots,a_k\}$). The resulting answer may  be  redundant, if the representative has no other adjacent edges in the answer other than equivalence edges. In such cases, 
in a \textbf{post-processing step}, we remove  from the answer the representative and its equivalence edges, then reconnect  the respective equivalent nodes using $k-1$ equivalence edges. This guarantees obtaining a non-redundant tree, such as the one in Figure~\ref{fig:prop-example}(c).

\subsection{The GAM algorithm}

\begin{figure*}[t!]
\fbox{
\begin{minipage}{\textwidth}
Procedure \textbf{process}(tree $t$)
\begin{itemize}[noitemsep,nolistsep]
\item if $t$ is not already in $E$
 \item then
\begin{itemize}[noitemsep,nolistsep]
\item add $t$ to $E$
\item if $t$ has matches for all the query keywords
\item then post-process $t$ if needed; output the result as an answer
\item else insert $t$ into $K$
\end{itemize}
\end{itemize}

Algorithm \textbf{GAMSearch}(query $Q=\{w_1,w_2,\ldots, w_k\}$)
\begin{enumerate}[noitemsep,nolistsep]
\item For each $w_i$, $1\leq i \leq k$
\begin{itemize}[noitemsep,nolistsep]
\item For each node $n_i^j$ matching $w_i$, let $t_i^j$ be the 1-node tree consisting of $n_i^j$; process($t_i^j$) \label{item:1node}
\end{itemize}
 \item Initial \textsc{merge}$^*$: try to merge every pair of trees from
   $E$, and process any resulting answer tree. \item Initialize $U$ (empty so far):  \label{item:init-Q}
\begin{enumerate}[noitemsep,nolistsep]
\item Create \grow\ opportunities: Insert into $U$ the pair $(t,e)$, for each $t\in E$ and $e$ a data or similarity edge adjacent
   to $t$'s root.
\item Create \gtr\ opportunities: Insert into $U$ the pair $(t, n\rightarrow n_{rep})$ for each $t\in E$ whose root is $n$, such that the representative of $n$ is $n_{rep}\neq n$.
\end{enumerate}
 \item While ($U$ is not empty)
 \begin{enumerate}[noitemsep]
 \item Pop out of $U$ the highest-priority pair $(t,e)$.
\item Apply the corresponding \grow\ or \gtr, resulting in a new tree $t''$;  process($t''$).
\item If $t''$ was not already in $E$, agressively \mergecur:
\begin{enumerate}[noitemsep,nolistsep]
\item Let $NT$ be a set of new trees obtained from the \mergecur\ (initially $\emptyset$).
\item Let $\mathbf{p_1}$ be the keyword set of $t''$
\item For each  keyword subset $\mathbf{p_2}$ that is a
  key within $K$, and such that $\mathbf{p_1}\,\cap\, \mathbf{p_2} =\emptyset$
\begin{enumerate}[noitemsep,nolistsep]
\item For each tree $t^i$ that corresponds to $\mathbf{p_2}$, try to merge $t''$ with $t^i$. Process any possible result; if it is new (not in $E$ previously), add it to $NT$. \label{item:merge-candidates}
\end{enumerate}
\end{enumerate}
\item Re-plenish $U$ (add more entries in it) as in step~\ref{item:init-Q}, based on the trees $\{t''\} \, \cup \, NT$. \label{item:replenish-Q}
\end{enumerate}
\end{enumerate}
\end{minipage}}
\caption{Outline of GAM algorithm\label{fig:algo}}
\end{figure*}

We now have the basic exploration steps we need: \grow, \gtr\ and \mergecur.
In this section, we explain how we use them in our integrated keyword search algorithm.

We decide to apply in sequence: one \grow\ or \gtr\ (see below), leading to a new tree $t$, immediately followed by all the \mergecur\ operations possible on $t$. Thus, we call our algorithm \textbf{Grow and Aggressive Merge} (GAM, in short). We merge aggressively in order to detect as quickly as possible when some of our trees, merged at the root, form an answer.

Given that every node of a currently explored answer tree can be connected with several edges, we need to decide which \grow\ (or \gtr) to apply at a certain point. For that, we use a  \textbf{priority queue} $U$ in which we add (tree, edge) entries: for \grow, with the notation above, we add the $(t, e)$ pair, while for \gtr, we add $t$ together with the equivalence edge leading to the representative of $t$'s root. In both cases, when a $(t, e)$ pair is extracted from $U$, we just extend $t$ with the edge $e$ (adjacent to its root), leading to a new tree $t_G$, whose root is the other end of the edge $e$. Then we aggressively merge $t_G$ with all compatible trees explored so far, finally we read from the graph the (data, similarity or equivalence) edges adjacent to $t_G$'s root and add to $U$ more (tree, edge) pairs to be considered further during the search. The algorithm then picks the highest-priority pair in $U$ and reiterates; it stops when $U$ is empty, at a timeout, or when a maximum number of answers are found (whichever comes first).

The last parameter impacting the exploration order is the priority used in $U$: at any point, $U$ gives the highest-priority $(t, e)$ pair, which determines the operations performed next.
\begin{enumerate}
\item Trees matching \emph{many query keywords} are preferable, to go toward complete query answers;
\item At the same number of matched keywords, \emph{smaller trees} are preferable in order not to miss small answers;
\item Finally, among $(t_1,e_1)$, $(t_2,e_2)$ with the same number of nodes and matched keywords,  we prefer the pair with the \emph{higher specificity edge}. 
\end{enumerate}


\noindent\textbf{Algorithm details} Beyond the priority queue $U$ described above, the algorithm also uses a {\em memory of all the trees explored}, called $E$. It also organizes all the (non-answer) trees into a map $K$ in which they can be accessed by the subset of query keywords that they match. The algorithm is shown in pseudocode in Figure~\ref{fig:algo}, following the notations introduced in the above discussion.

While not  shown in Figure~\ref{fig:algo} to avoid clutter, the algorithm {\em only develops minimal trees} (thus, it only finds minimal answers). This is guaranteed:
\begin{itemize}
\item When creating \grow\ and \gtr\ opportunities (steps \ref{item:init-Q} and \ref{item:replenish-Q}): we check not only that the newly added does not close a cycle, but also that the matches present in the new tree satisfy our minimality condition (Section~\ref{sec:search-problem}).
\item Similarly, when selecting potential \mergecur\ candidates (step \ref{item:merge-candidates}).
\end{itemize}

\section{Experimental evaluation}
\label{sec:evaluation}

The algorithms described above are implemented in the \textbf{ConnectionLens} prototype, available
\href{https://gitlab.inria.fr/cedar/connectionlens}{online}. 
Below, we report the results of an experimental evaluation we carried out to study the performance of its algorithms, as well as quality aspects of the constructed graphs. 
Section~\ref{sec:evaluation:setup} describes the software and hardware setup, and our datasets. Section~\ref{sec:evaluation:construction} focuses on graph construction, while Section~\ref{sec:evaluation:query} targets keyword query answering. 

\subsection{Software, hardware, and datasets}
\label{sec:evaluation:setup}

ConnectionLens is a \textbf{Java application} (44.700 lines) which relies on a relational database to store the constructed graphs and as a back-end used by the query algorithm. It features controllable-size caches to keep in memory as many nodes and edges as possible; this allows adapting to different memory sizes.
It also comprises \textbf{Python} code (6.300 lines) which implements entity extraction (Section~\ref{sec:entity-extraction}) and content extraction from PDF documents to JSON (see~\cite{construction-paper}), tasks for which the most suitable libraries are in Python. 
The Flair extractor (Section~\ref{sec:entity-extraction}) and the disambiguator (Section~\ref{sec:disambig}) are Web services which ConnectionLens calls. The former is deployed on the machine where ConnectionLens runs. We deployed the latter on a dedicated Inria server,  adapting the original Ambiverse code to our new pipeline introduced in Section~\ref{sec:disambig}; the disambiguator  consists of $842$ Java classes. 

For our experiments, we used a \textbf{regular server} from 2016, equipped with 2x10-core Intel Xeon E5-2640 (Broadwell) CPUs clocked at 2.40GHz, and 128GB DRAM, which uses PostgreSQL 12.4 to store the graph content in a set of tables.
This is a medium-capacity machine, without special capabilities; recall our requirement \textbf{R3} that our algorithms be feasible on off-the-shelf hardware.
We also used a \textbf{GPU server} from 2020, with a 2x16-core Intel Xeon Gold 5218 (Skylake) CPUs clocked at 2.30GHz, an NVIDIA Tesla V100 GPU and 128GB DRAM.
To show the applicability of our software to standard hardware configurations, we focus on the results that we obtained with our regular server.
However, we also include some results on the more advanced server to show that our platform adapts seamlessly to modern as well as heterogeneous hardware, which includes both CPUs and GPUs.
When needed to separate them, we will refer to each server with its CPU generation name.

\noindent\textbf{Data sources} Most of our evaluation is on {\em real-world} datasets, described below from the smallest to the largest (measuring their size on disk before being input to ConnectionLens).
\noindent\textbf{1.} We crawled the French online newspaper Mediapart and obtained 256 articles for the search keywords ``{\em corona, chloroquine, covid}'' (256 documents), and 886  for ``economie, chomage, crise, budget'' (1142 documents and \textbf{18.4 MB} overall). 
\noindent\textbf{2.} An \textbf{XML document}\footnote{https://www.hatvp.fr/livraison/merge/declarations.xml} comprising business interest statements of French public figures, provided by  HATVP ({\em Haute Autorit\'{e} pour la Transparence de la Vie Publique}); the file occupies \textbf{35 MB}. 
\noindent\textbf{3.} A subset of the \textbf{YAGO 4}~\cite{yago4} RDF knowledge base, comprising entities present in the French Wikipedia and their properties; this takes \textbf{2.49 GB} on disk (17.36 M triples). 
For a  fine-granularity, controlled study of our query algorithm, we also devised a set of {\em small synthetic graphs}, described 
in Section~\ref{sec:evaluation:query}.

\subsection{Construction evaluation}
\label{sec:evaluation:construction}

We present results of our experiments measuring the performance and the quality of the modules involved in graph construction.
We study  graph construction performance in
Section~\ref{sec:exp-construction}, the quality of our
information extraction in Section~\ref{sec:exp-ner}, and that of 
disambiguation in Section~\ref{sec:exp-disambiguation}.

\subsubsection{Graph construction}
\label{sec:exp-construction}

We start by studying the impact of \textbf{node factorization} (Section~\ref{sec:graph-refine-optim}) on the number of graph nodes and the graph storage time. For that, we rely on the XML dataset, and {\em disable entity extraction, entity disambiguation, and node matching}. Its (unchanged) number of edges $|E|$ is  $1.588.839$.
For each type of loading, we report  the number of nodes $|N|$, the time spent storing nodes and edges to disk $T_{DB}$, and the total running time $T$ in Table~\ref{tab:impact}.

\begin{wrapfigure}{L}{0.62\textwidth}
\vspace{-3.5mm}
\scalebox{0.9}{\begin{tabular}{p{3.7cm}rrrrr}
Value node creation policy &  $|N|$ &  $T_{DB}$ (s) & $T$ (s)\\\hline
Per-instance &  $1.588.840$ &  $318$ & $319$ \\
Per-path & $1.093.060$  & $271$& $318$ \\
Per-path w/ null code detection & $1.207.951$ & $276$ & $323$ \\
Per-dataset & $1.084.918$ &  $265$ & $307$\\
Per-graph & $1.084.918$ & $179$ & $228$ \\
Per-graph w/ null code detection & $1.199.966$ & $179$ & $229$ \\
\end{tabular}}
\vspace{-2.5mm}
\caption{Impact of node factorization.} 
\label{tab:impact}
\vspace{-3mm}
\end{wrapfigure}


Moving from per-instance to per-path node creation reduces the number of nodes by a third. However, this introduces some errors, as the dataset features many \textbf{null codes} (Section~\ref{sec:graph-refine-optim}); for instance, with per-instance value creation, there are 
$1.113$ nodes labeled {\em ``n\'{e}ant''} (meaning ``non-existent''),
$32.607$ nodes labeled {\em ``Donn\'{e}es non publi\'{e}es''} (unavailable information) etc. Using per-path, the latter are reduced to just $1.154$, which means that in the dataset, {\em ``Donn\'{e}es non publi\'{e}es''} appears $32.607$ times on $1.154$ different paths. However, this factorization, which introduces  connections between the XML nodes which are parents of this ``null'' value, may be seen as wrong. When the null codes were input to ConnectionLens, such nodes are no longer unified; the number of nodes increases, and so does the storage time.
In this graph, consisting of one data source, per-dataset and per-graph produce the same number of nodes, overall the smallest; it also increases when null codes are not factorized.
We conclude that {\em per-graph value creation combined with null code detection} is a practical alternative.

Next, we study the impact of \textbf{named entity extraction} (Section~\ref{sec:entity-extraction}) and \textbf{disambiguation} (Section~\ref{sec:disambig}) on the graph construction time.

\begin{wrapfigure}{L}{0.6\textwidth}
\includegraphics[width=0.6\textwidth]{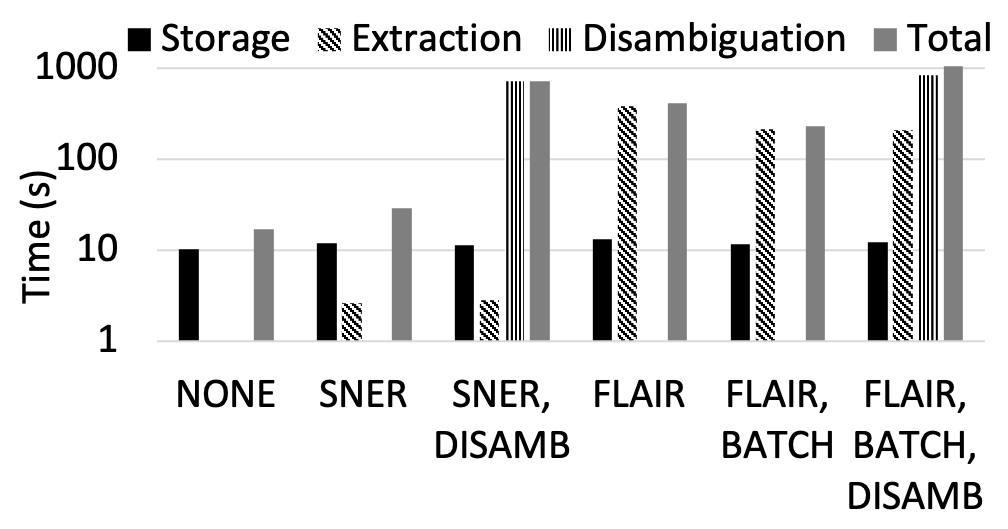}
\vspace{-10mm}
\caption{Graph construction time (seconds).\label{fig:impact-extract-disambig}}
\vspace{-5mm}
\end{wrapfigure}

For this, we load  $100.000$ triples from our Yago subset, with per-graph factorization, natural for the RDF data model where each literal or URI denotes one node in the RDF graph. In Figure~\ref{fig:impact-extract-disambig}, we load the triples using several configurations: without any entity extraction (NONE); with SNER entity extraction, without and then with disambiguation; with FLAIR entity extraction, without and then with disambiguation.
While Flair is slower, its results are qualitatively much better than those obtained with SNER (see Section~\ref{sec:exp-ner} below). To make it faster,
we also implement a \textbf{batch extraction} mechanism whereas $l_B$ labels are input a time to each extraction service, to take advantage of the parallel processing capacity available in current CPUs. In Figure~\ref{fig:impact-extract-disambig}, in the ``FLAIR, BATCH'' column, we used $l_B=128$ which maximized performance in our experiments.
A second optimization leverages  the fact that so-called {\em sequence to sequence (seq2seq)} models such as that used in our Flair extractor, when given a batch of inputs, pad the shortest ones to align them to the longest input, and some computation effort is lost on useless padding tokens. Instead, if {\em several batches}, say $n_B$, are received by the extraction service, it can {\em re-group} the inputs  so that one call to the seq2seq model is made over inputs of very similar length, thus no effort is wasted. In our experiments,  we used $n_B=10$.

Figure~\ref{fig:impact-extract-disambig} shows the time taken by storage, extraction and disambiguation (when applied) and the total graph creation time; note the logarithmic $y$ axis.  Storing the graph PostgreSQL dominates the loading time in the absence of extraction, or when using SNER. In contrast, Flair extraction takes more than one order of magnitude longer; batching reduces it by a factor of two. Finally, disambiguation, relying on computationally complex operations, takes longest;  it also incurs a modest invocation overhead as it resides on a different server (with the regular server hardware described in Section~\ref{sec:evaluation:setup}),  in the same fast network.

\noindent Next, we study \textbf{node matching} (Section~\ref{sec:matching}). For this, we loaded the XML dataset, which comes from individual declaration of interest, filled in by hundreds of users, with numerous spelling variations, small errors and typos. Loaded per-graph mode, with batched Flair extraction, the resulting graph has  $1.102.209$ nodes and $1.615.291$ edges.
We test two configurations: comparing {\em all leaf node pairs} to find possible similarity edges, respectively, comparing {\em only entity pairs}.
On the regular server, in both cases, data storage took $3$ minutes, and extraction $13$ minutes. When all comparisons are made, they take $39$ minutes, dominating the total time of $56$ minutes. A total of $28.875$ similar pairs are found to be above our similarity thresholds,
and lead to the same number of cl:sameAs edges stored in the graph, together with the respective similarity values. Only $748$ among these are entity pairs; the others are pairs of similar strings. This confirms the interest of node matching; we hope to reduce its cost further by using techniques such as Locality-Sensitive Hashing (LSH).


\begin{wrapfigure}{L}{0.6\textwidth}
\vspace{-4mm}
\includegraphics[width=0.6\textwidth]{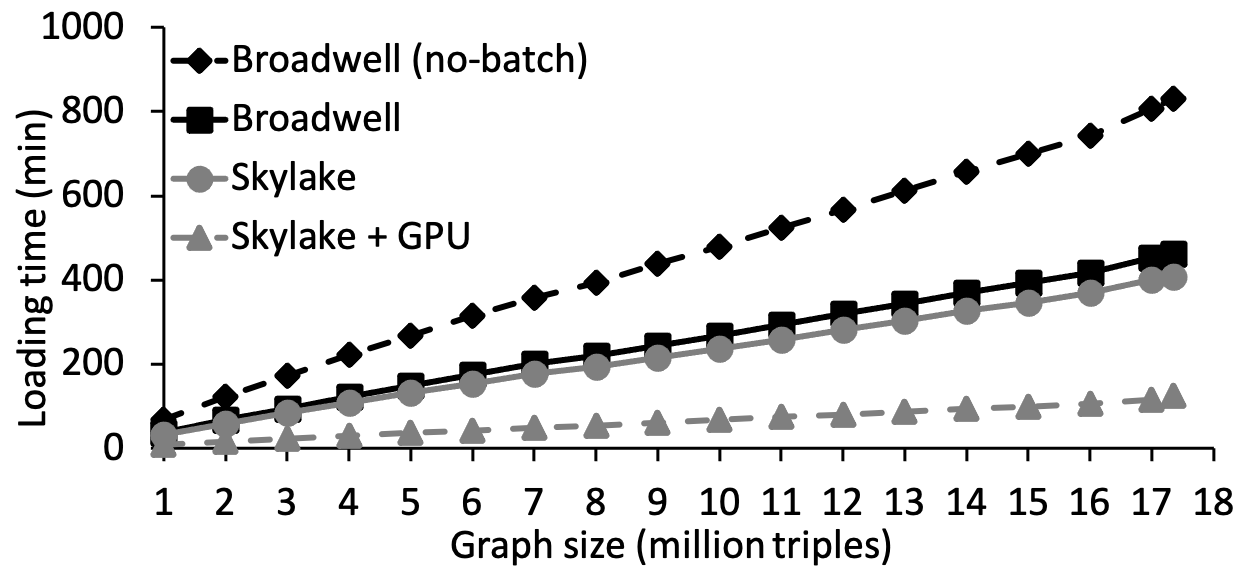}
\vspace{-8mm}
\caption{YAGO loading time (minutes) using Flair. \label{fig:yago}}
\vspace{-4mm}
\end{wrapfigure}

To study the \textbf{scalability} of our loading process, we loaded our YAGO subset by slices of 1M triples and measured the running time for these increasing data sizes, using the best extractor (Flair) with the same batch size(s) as above.  Figure~\ref{fig:yago} shows the loading time as the data grows, in three different hardware settings: on our regular server, our more powerful GPU server {\em disabling GPU use}, and the same {\em exploiting the GPU}.
Figure~\ref{fig:yago} shows that loading time scales linearly in the data volume on all configurations, and batching helps make the most out of our regular server.

\noindent \textbf{Complete graph} Finally, we  loaded all the data sources described in Section~\ref{sec:evaluation:setup} in a single graph, using per-graph node creation, batched Flair extraction, and disambiguation for HTML and XML (not for the RDF Yago subset, whose literals are already associated to URIs).
The graph has $|N|=8.019.651$ nodes (including $677.459$ person entities, $275.316$ location entities, and $61.452$ organization entities), and $|E|=20.642.207$ edges. Many entities occur across data sources, e.g., $330$ person entities occur, each, in at least $10$ sources; the French president E.~Macron occurs in $183$ distinct sources. All these lead to interconnections across sources. On our fastest (GPU) hardware configuration, loading the RDF data took $128$ minutes; the HTML articles another $260$, and the XML document $96$ minutes more. The last two times reflect the relatively high cost of  disambiguation (recall  Figure~\ref{fig:impact-extract-disambig}). ConnectionLens users can turn it off when it is not needed (e.g., if users feel they know the real-world entities behind entity labels encountered in the graph), or trigger it {\em selectively}, e.g., on organizations but not on people nor locations etc.

\subsubsection{Named-Entity Recognition quality}
\label{sec:exp-ner}

Due to the unavailability of an off-the-shelf, good-quality entity extractor for French text, we decided to train a new model.
To decide the best NLP framework to use, we experimented with the Flair~\cite{akbik2019flair} and SpaCy (\url{https://spacy.io/})  frameworks. Flair allows {\em combining} several embeddings, which can lead to significant quality gains. Following~\cite{akbik2019flair}, after testing different word embedding configurations, 
we trained a Flair model using {\em stacked forward and backward French Flair embeddings} with {\em French fastText embeddings} on the WikiNER dataset. We will refer to this model as \textit{Flair-SFTF}.


Below, we describe a \emph{qualitative comparison} of \textit{Flair-SFTF}  with the French Flair and SpaCy {\em pre-trained} models.
The French pre-trained Flair model 
is trained with the WikiNER dataset, and uses French character embeddings trained on Wikipedia, and French fastText embeddings.
As for SpaCy, two pre-trained models are available for French: a medium (\textit{SpaCy-md}) and a small one (\textit{SpaCy-sm}). They are both trained with the WikiNER dataset and the same parameterization. The difference is that \textit{SpaCy-sm} does not include word vectors, thus, in general,  \textit{SpaCy-md}  is expected to perform better, since word vectors will most likely impact positively the model performance. 
Our evaluation also includes the model previously present in ConnectionLens~\cite{Chanial2018}, trained using \textit{Stanford NER}~\cite{finkel2005incorporating}, with the Quaero Old Press Extended Named Entity corpus~\cite{galibert2012extended}.

We measured the precision, recall, and $F1$-score of each model using the \textit{conlleval} evaluation script, previously used for such tasks\footnote{The script \url{https://www.clips.uantwerpen.be/conll2002/ner/} has been made available in conjunction with the CoNLL (Conference on Natural Language Learning).}. \textit{conlleval} evaluates \emph{exact matches}, i.e.,  both the text segment of the proposed entity and its type, need to match ``gold standard'' annotation, 
 to be considered correct. Precision, recall, and $F1$-score (harmonic mean of precision and recall) are computed for each named-entity type.
To get an aggregated, single quality measure, \textit{conlleval} computes the {\em micro-average} precision, recall, and $F1$-score over all recognized entity instances, of all named-entity types.

For evaluation, we used the entire FTBNER dataset~\cite{sagot-etal-2012-annotation}.
We pre-processed it to convert its entities from the seven types they used, to the three we consider, namely, persons, locations and organizations.
After pre-processing, the dataset contains $12$K sentences 
and $11$K named-entities ($2$K persons, $3$K locations and $5$K organizations).

\begin{table}[h!]
\begin{tabular}{|p{0.14\columnwidth}|p{0.12\columnwidth}|p{0.14\columnwidth}|p{0.14\columnwidth}|p{0.14\columnwidth}|p{0.14\columnwidth}|}\cline{1-6}
Entities & Flair-SFTF & Flair-pre-trained & SpaCy-md & SpaCy-sm & Stanford NER \\\hline
LOC-P       & 59.52\% & 53.26\% & 55.77\% & 54.92\% & 62.17\% \\
LOC-R       & 79.36\% & 77.71\% & 78.00\% & 79.41\% & 69.05\% \\
LOC-$F1$    & 68.02\% & 63.20\% & 65.04\% & 64.93\% & 65.43\% \\\hline
ORG-P       & 76.56\% & 74.57\% & 72.72\% & 71.92\% & 15.82\% \\
ORG-R       & 74.55\% & 75.61\% & 54.85\% & 53.23\% & 5.39\%  \\
ORG-$F1$    & 75.54\% & 75.09\% & 62.53\% & 61.18\% & 8.04\%  \\\hline
PER-P       & 72.29\% & 71.76\% & 53.09\% & 57.32\% & 55.31\% \\
PER-R       & 84.94\% & 84.89\% & 74.98\% & 79.19\% & 88.26\% \\
PER-$F1$    & 78.10\% & 77.78\% & 62.16\% & 66.50\% & 68.00\% \\\hline
Micro-P     & 69.20\% & 65.55\% & 61.06\% & 61.25\% & 50.12\% \\
Micro-R     & 77.94\% & 77.92\% & 65.93\% & 66.32\% & 40.69\% \\
Micro-$F1$  & 73.31\% & 71.20\% & 63.40\% & 63.68\% & 44.91\% \\\hline
\end{tabular}
\caption{Quality of  NER from French text.\label{fig:ner-results-table}}
\vspace{-4mm}
\end{table}

The evaluation results are shown in Table \ref{fig:ner-results-table}.
All models perform better overall than the \textit{Stanford NER} model previously used in ConnectionLens~\cite{Chanial2018}, which has a micro $F1$-score of about 45\%. 
The \textit{SpaCy-sm} model has a slightly better overall performance than \textit{SpaCy-md}, with a small micro $F1$-score difference of $0.28\%$. \textit{SpaCy-md} shows higher $F1$-scores for locations and organizations, but is worse on people, driving down its overall quality. 
All Flair models surpass the micro scores of SpaCy models. In particular, for people and organizations, Flair models show more than $11\%$ higher $F1$-scores than SpaCy models. Flair models score better on all named-entity types, except for locations when comparing the SpaCy models, specifically, with the \textit{Flair-pre-trained}.
\textit{Flair-SFTF} has an overall $F1$-score of $73.31\%$ and has better scores than the \textit{Flair-pre-trained} for all metrics and named-entity types, with the exception of the recall of organizations, lower by $1.06\%$.
In conclusion, {\em Flair-SFTF} is the best NER model we evaluated.

\subsubsection{Disambiguation quality}
\label{sec:exp-disambiguation}

We now  evaluation the quality of the disambiguation module.
As mentioned in Section~\ref{sec:disambig}, our module works for both English and French.

\begin{wrapfigure}{L}{0.55\textwidth}
\begin{tabular}{|l|c|c|c|}\cline{2-4}
\multicolumn{1}{l|}{}
        & Precision &  Recall  & $F1$\\\hline
LOC     &   99.00\% &  97.05\% &  98.01\% \\
ORG     &   92.38\% &  75.19\% &  82.90\% \\
PER     &   75.36\% &  77.94\% &  76.62\% \\\hline
Micro &   90.51\% &  82.94\% &  86.55\% \\\hline
\end{tabular}
\caption{Quality of disambiguation for French.\label{fig:ned}}
\vspace{-4mm}
\end{wrapfigure}

The performance for English has been measured on the CoNLL-YAGO dataset~\cite{hoffart2011robust}, by the developers of Ambiverse. They report a micro-accuracy of $84.61\%$ and a macro-accuracy of $82.67\%$.
To the best of our knowledge, there is no labeled corpus for entity disambiguation in French, thus we evaluate the performance of the module on the FTBNER dataset previously introduced.
FTBNER consists of sentences annotated with named entities. The disambiguation module takes a sentence, the type, and offsets of the entities extracted from it, and returns for each entity either the URI of the matched entity or an empty result if the entity was not found in the KB.  In our experiment,  $19\%$ of entities have not been disambiguated, more precisely $22\%$ of organizations, $29\%$ of persons, and $2\%$ of locations.
For a fine-grained error analysis, we sampled 150 sentences and we manually verified the disambiguation results (Figure~\ref{fig:ned}). The module performs very well, with excellent results for locations ($F1 = 98.01\%$), followed by good results for organizations ($F1 = 82.90\%$) and for persons ($F1 = 76.62\%$). In addition to these results, we obtain a micro-accuracy of $90.62\%$ and a macro-accuracy of $90.92\%$. The performance is comparable with the one reported by the Ambiverse authors for English. We should note though that the improvement for French might be due to our smaller test set.



\subsection{Answering keyword queries}
\label{sec:evaluation:query}

This section presents the results that we obtained by using synthetic and real-world datasets.
In each case, we first describe the datasets, and then we present and explain our findings.
We bound the query execution time to 120 seconds, after which the algorithm stops searching for matches.

\subsubsection{Queries on synthetic datasets}
We first study the performance of our algorithm on  different types of synthetic datasets.
The first type is the \emph{line graph}, where every node is connected with two others, having one edge for each node, except two nodes which are connected with only one.
We use the line graph to clearly show the performance of \grow~and \mergecur~operations with respect to the graph size.
The second type is the \emph{chain graph}, which is the same as the line, but it has two edges (instead of one) connecting every pair of nodes.
We use the chain graph to show the performance of the algorithm as we double the amount of edges of the line graph and we give more options to \grow~and \mergecur.
The third type is the \emph{star graph}, where we have several line graphs connected through a strongly connected cluster of nodes with a representative.
We use this type to show the performance of \gtr, by placing the query keywords on different line graphs.
The fourth type is a random graph based on the \emph{Barabasi-Albert}
(BA) model~\cite{Barabasi509}, which generates scale-free networks with only a few nodes (hubs) of the graph having much higher degree than the rest.
The graph in this model is created in a two-staged process.
During the first stage, a network of some nodes is created.
Then, during the second stage, new nodes are inserted in the graph and they are connected to nodes created during the first stage.
We set every node created at the second stage to be connected with exactly one node created at the first stage.
In the following, The black line shows the time elapsed until the first answer is found, whereas the grey line shows the overall execution time.




\begin{figure*}[t!]
\centering
\subfloat[Line graph\label{fig:eval/line}]{
  \begin{minipage}{0.45\columnwidth}
    \includegraphics[width=\columnwidth]{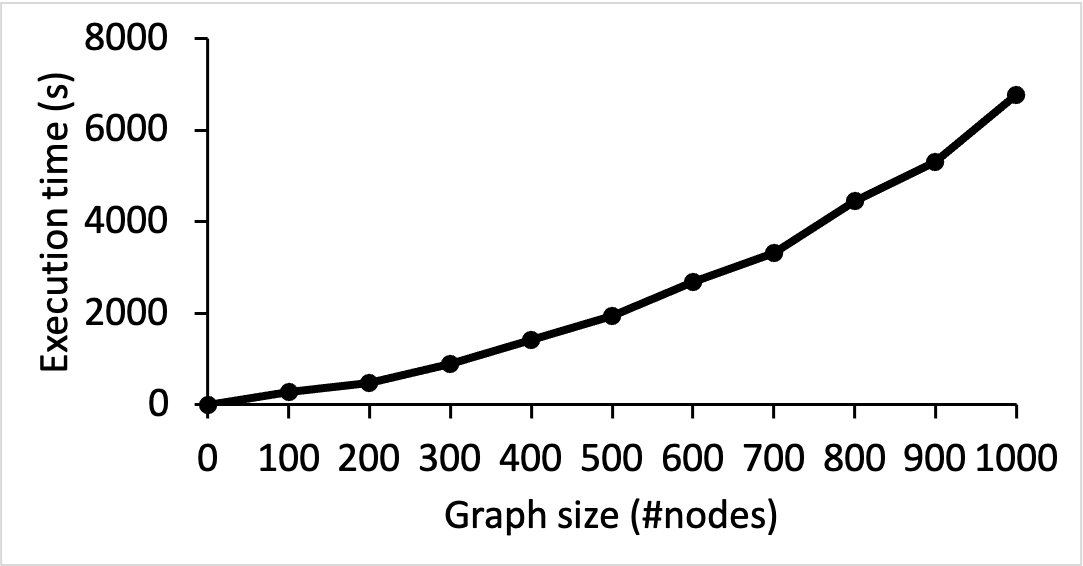}
  \end{minipage}
}
\quad
\subfloat[Chain graph\label{fig:eval/chain}] {
  \begin{minipage}{0.45\columnwidth}
    \includegraphics[width=\columnwidth]{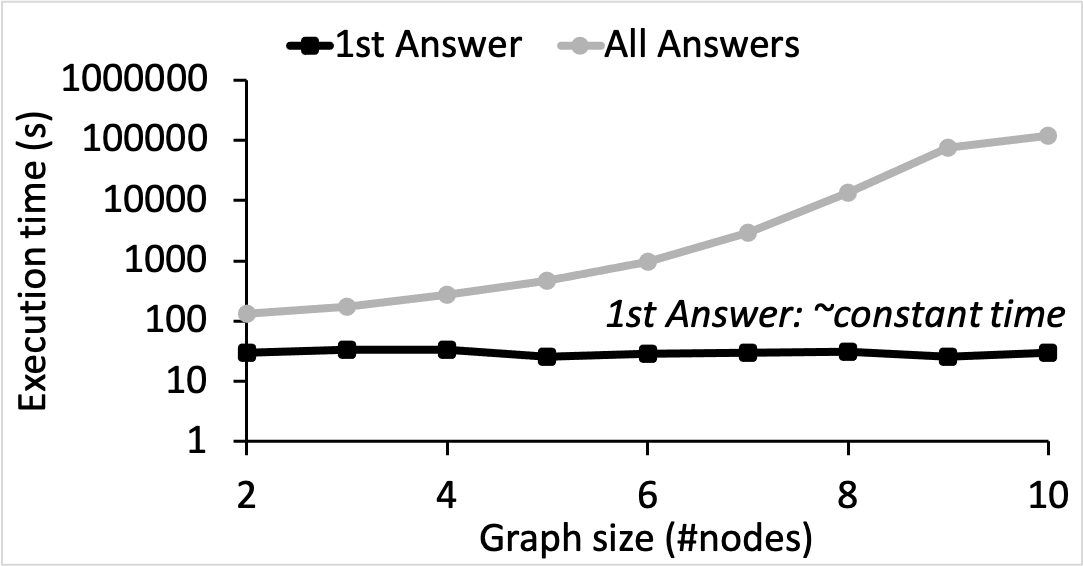}
  \end{minipage}
}
\vspace{-4mm}
\subfloat[Star graph\label{fig:eval/star}] {
  \begin{minipage}{0.45\columnwidth}
    \includegraphics[width=\columnwidth]{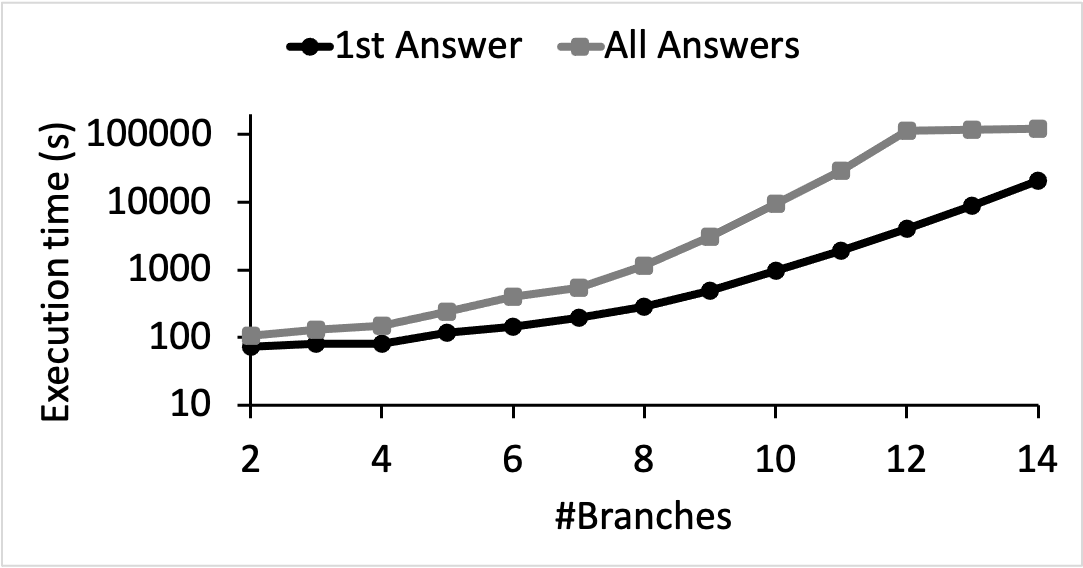}
  \end{minipage}
}
\quad
\subfloat[Barabasi-Albert\label{fig:eval/ba}] {
  \begin{minipage}{0.45\columnwidth}
    \includegraphics[width=\columnwidth]{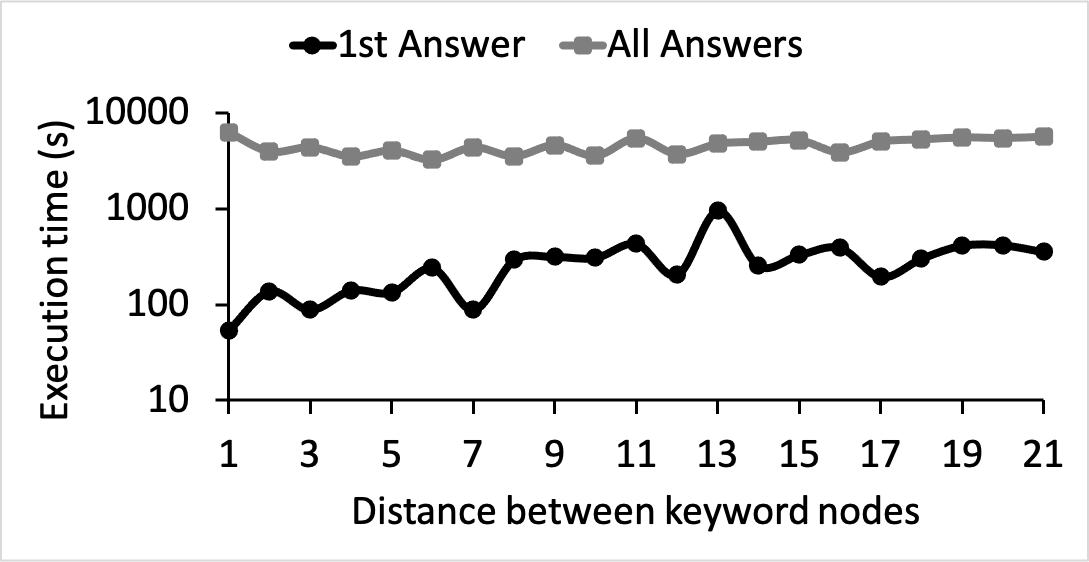}
  \end{minipage}
}

\vspace{-4mm}
\caption{Query execution time on different synthetic graph types.}
\label{fig:eval/synthetic}
\vspace{-6mm}
\end{figure*}

Figure~\ref{fig:eval/synthetic} includes the results that we obtained by querying the synthetic datasets.
The black line shows the time elapsed until the first answer is found, whereas the grey line shows the overall execution time.

Figure~\ref{fig:eval/line} shows the execution time of our algorithm when executing a query with two keywords on a line graph, as we vary the number of nodes of the graph.
We place the keywords on the two ``ends'' of the graph to show the impact of the distance on the execution time.
The performance of our algorithm is naturally affected by the size of the graph, as it generates $2\cdot N$ answer trees, where $N$ is the number of nodes.
Given that this is a line graph, there is only one answer, which is the whole graph, and, therefore, the time to find the first answer is also the overall execution time.


Figure~\ref{fig:eval/chain} shows the performance of our algorithm on a chain graph.
The execution times for the first answer are almost the same, as the graph size increases slowly.
Instead, the overall execution times increase at a much
higher (exponential) rate; note the logarithmic scale of the $y$ axis.
The reason is that every pair of nodes is connected with two edges, which increases the amount of answers exponentially with the amount of nodes in the graph.


In Figure~\ref{fig:eval/star}, we report the execution time for the star graph.
We place keywords in two different lines connected through the center of the graph, forcing the algorithm to use \gtr, whereas in the previous cases it only had to use \grow~and \mergecur.
The number of branches, depicted on the $x$ axis, corresponds to the number of line graphs connected in the star.
Each line graph has 10 nodes and we place the query keywords at the extremities of two different line graphs.
The number of merges is exponential to the number of branches, that is $\mathcal{O}(2^K)$ where $K$ is the number of branches, since the algorithm will check all possible answers.
This behaviour is clearly shown in both lines of
Figure~\ref{fig:eval/star}, where on the $y$ axis (in logarithmic
scale) we show the times to find the first, and, respectively, all
answers.
Above 12 branches, the timeout of 120 seconds that we have set is hit and, thus, search is terminated, as shown when we search for all answers.


Figure~\ref{fig:eval/ba} depicts the query performance with the Barabasi-Albert model.
We fix the graph size to 2000 nodes and we vary the position of two keywords, by choosing nodes which have a
distance, as given in the $x$ axis; note the logarithmic $y$ axis.
As the graph is randomly generated within the BA model, we note some irregularity in the time to the first solution, which however grows at a moderate pace as the distance between the keyword node grows.
The overall relation between the time to the first solution and the total time confirms that the search space is very large but that most of the exploration
is not needed, since the first solution is found quite fast.

\vspace{-1.5mm}
\subsubsection{Queries on the complete, real-world graph}
\vspace{-1mm}

\begin{table*}[h!]
 \begin{adjustbox}{width=\textwidth}
  \centering
  \begin{tabular}{ |c|r|r|r| }
    \hline
    Query keyword(s)                                                                        & Answers & Answer trees & Time to 1st (ms) \\ 
    \hline\hline
{\em a\'eronautique}, {\em Macron}             &        2779  &  1152577 &  7225 \\ 
\hline
Brigitte Macron, Clara Gaymard                  &       4584 &          545020 &       1412        \\ 
\hline
{\em Chine}, {\em covid}, {\em France}                 &              17           &         25974     & 8380 \\ 
\hline
{\em ch\^omage}, {\em covid}                      &       108 & 205584 & 4476 \\
\hline
{\em covid}, El Khomri                                                & 16 & 215952 & 52486 \\ 
\hline
{\em confinement}, Christophe Castaner                    &  4 & 120367& 3820  \\ 
\hline
Chine, France, Didier Raoult       &            36 &         146261 &     14666      \\ 
\hline
Ebola, Raoult                                               &              1 &           37751 &      75759    \\ 
\hline
{\em entreprise}, {\em Raffarin}                   &        6336 & 1174822 & 6589 \\
\hline
Julien Denormandie, Macron                         &        464    &          20181  &  2661        \\ 
\hline
Khalid al-Falih, Kristalina Georgieva                &            1    &            1775 &     765        \\ 
\hline
Kristalina Georgieva, Walter Butler                     &             1   &   3224         &           353      \\ 
\hline
Louis Beam, Ku Klux Klan, Trump                     &             1     & 24172        &          15207    \\ 
\hline
{\em Macron}, {\em Royal}                                  &     102  & 6413                &  4107 \\
\hline
Marisol Touraine, Jean-Fran\c{c}ois Delfraissy &    3            & 1497     & 1183 \\ 
\hline
{\em masque}, {\em France}                              &    35     &  15082  &  7126 \\
\hline
Michael Ryan, Anthony Fauci                           &              2 &  6653            &            12215     \\ 
\hline
Pascale Gruny, Jean-Fran\c{c}ois Delfraissy             & 2 & 1332       &   91591  \\ 
\hline
{\em vaccination}, {\em Trump} & 12 & 1188 & 6518 \\ 
\hline
Yazdan Yazdanpanah, Jean-Fran\c{c}ois Delfraissy & 29 & 5446 & 1687  \\ 

    \hline
  \end{tabular}
  \end{adjustbox}\vspace{-1.5mm}
  \caption{Query results on the complete graph.}
  \label{tbl:eval/realworld}
\vspace{-5mm}
\end{table*}

Next, we describe results  that we obtained querying a graph obtained by loading all the real-world data sources decribed in Section~\ref{sec:evaluation:setup}.
We report our findings in Table~\ref{tbl:eval/realworld}. The queries feature terms that appear in recent French news; they are related to the economy, the Covid crisis, world events, and/or French politics. In most queries, keywords are exact entity names, with their most common spelling. In other cases, we allowed node labels to {\em approximately} match a keyword (based on PostgreSQL' stemming and string pattern matching); such keywords are shown in italic in the table. Again, we gave a timeout of 2 minutes, and on this large graph, all the GAM searches stopped at a time-out. The table shows that queries return varied number of answers, but many ATs are developed in all cases, and the first is found quite before the timeout. 
Finally, an inspection of the results showed that most are obtained from different datasets, confirming the interest of linking datasets in ConnectionLens. These results show the feasibility and interest of GAMSearch on large heterogeneous graphs. 

\mysection{Related work and perspectives}
\label{sec:related}


Our work belongs to the area of {\em data integration}~\cite{doan2012int}.
Data integration can be achieved either in a {\em warehouse} fashion (consolidating all the data sources into a single repository), or in a {\em mediator} fashion (preserving the data in their original repository, and querying through a mediator module which distributes the work to each data source, and combines the results).
Initially, we followed a mediator approach~\cite{DBLP:journals/pvldb/BonaqueCCGLMMRT16}, allowing users to express queries using a mix of languages fitting every individual data source, similar to polystore systems~\cite{kolev2016cloudmdsql,alotaibi:hal-02070827}.
Our apporach could be further combined with a mixed XML-RDF language for fact-checking applications~\cite{goasdoue:hal-00814285,goasdoue:hal-00828906}.
However, the journalists' feedback was that the installation and maintenance of a mediator over several data sources, and querying through a mixed language, were very far from their technical skills.
This is why here, we ($i$)~pursue a {\bf warehouse} approach; ($ii$)~base our architecture on {\bf Postgres}, a highly popular and robust system; ($iii$)~simplify the query paradigm to keyword querying.



ConnectionLens integrates a wide variety of data sources: JSON, relational, RDF and text since~\cite{Chanial2018}, to which we have added XML, multidimensional tables, and  PDF documents.
We integrate such heterogeneous content in a graph, therefore, our work recalls the production of {\em Linked Data}.
A significant difference is that {\em we do not impose that our graph is RDF}, and {\em we do not assume, require, or use a domain ontology}.


Graphs are also produced when {\em constructing knowledge bases}, e.g., Yago~\cite{DBLP:conf/cidr/MahdisoltaniBS15,yago4}.
Our setting is more limited in that we are only allowed to integrate a given set of datasets that journalists trust to not ``pollute'' the database.
Therefore, we {\em use\/} a KB only for disambiguation and accept (as stated in Section~\ref{sec:intro}) that the KB does not cover some entities found in our input datasets. Our simple techniques for matching (thus, connecting) nodes are reminiscent of data cleaning, entity resolution~\cite{DBLP:journals/pvldb/SuchanekAS11,DBLP:conf/edbt/0001IP20}, and key finding in knowledge bases, e.g.~\cite{DBLP:journals/kbs/SymeonidouAP20}. Much more elaborate techniques exist, notably, when the data is regular, its structure is known and fixed, an ontology is available, etc.; none of these holds in our setting. 

\textit{Keyword search (KS)} is widely used for searching in unstructured (typically text) data, and it is also the best search method for novice users, as witnessed by the enormous success of keyword-based search engines.
As databases grow in size and complexity, KS has been proposed as a method for searching {\em also} in structured data~\cite{KS-book}, when users are not perfectly familiar with the data, or to get answers enabled by different tuple connections.
For relational data, in~\cite{DBLP:conf/vldb/HristidisP02} and subsequent works, tuples are represented as nodes, and two tuples are interconnected  through primary key-foreign key pairs.
The resulting graphs are thus quite uniform, e.g., they consist of ``Company nodes'', ``Employee nodes'' etc.
The same model was considered in~\cite{DBLP:conf/icde/OliveiraSM15,DBLP:conf/icde/SayyadianLDG07,DBLP:conf/sigmod/VuOPT08,DBLP:journals/debu/YuQC10}; \cite{DBLP:conf/icde/SayyadianLDG07} also establishes links based on similarity of constants appearing in different relational attributes.
As explained in Section~\ref{sec:steiner}, our problem is (much) harder since our trees can traverse edges in both directions, and paths can be (much) longer than those based on PK-FK alone.

KS has also been studied in \textbf{XML documents}~\cite{guo2003xrank,liu2007identifying}, where an answer is defined as a subtree of the original document, whose leaves match the query keywords. This problem is much easier than ours, since: ($i$)~an XML document is a tree, guaranteeing just one connection between any two nodes; in contrast, there can be any number of such connections in our graphs; ($ii$)~the maximum size of an answer to a $k$-keywords query is $k\cdot h$ where $h$, the height of an XML tree, is almost always quite small, e.g., $20$ is considered ``quite high''; in contrast, with our bi-directional search, the bound is $k\cdot D$ where $D$ is the diameter of our graph - which can be much larger.

Our \grow\ and \mergecur\ steps are borrowed from \cite{dpbf,blinks}, which address KS for \textbf{graphs}, assuming optimal-substructure, which does not hold for us, and single-direction edge traversal.  For \textbf{RDF graphs}~\cite{Elbassuoni:2011:KSO:2063576.2063615,DBLP:journals/tkde/LeLKD14} traverse edges in their direction only; moreover, \cite{DBLP:journals/tkde/LeLKD14} also make strong assumptions on the graph, e.g., that all non-leaf nodes have types, and that there are very few types (regular graph). In~\cite{error-tolerant}, the authors investigate a different kind of answers, called $r$-clique graphs, which they find using specific indexes.

Keyword search across \textbf{heterogeneous datasets} has been previously studied in~\cite{Dong:2007,Li:2008:EEK:1376616.1376706}. However, in these works, {\em each answer comes from a single dataset}, that is, they never consider answers spanning over and combining multiple datasets, such as the one shown in Figure~\ref{fig:example-graph}.
In turn, multiple datasets lead to equivalence and similarity edges; we have shown how to compactly encode the latter using representatives, and efficiently enumerate solutions which may span data, equivalence, and similarity edges.

A recent work \cite{DBLP:conf/sigir/LuPRAWW19} leverages text data sources to find more answers  to queries over knowledge graphs.  
An important  difference is that we integrate all datasets {\em prior to querying} whereas in \cite{DBLP:conf/sigir/LuPRAWW19}, text is accessed when required to complement the loaded data graph.
The effort we invest in building the graphs pays off by making queries faster.
Further, our work is more general in the data formats we support. Finally, \cite{DBLP:conf/sigir/LuPRAWW19} applies a set of heuristics, tied to the nature of the graphs they used, to find the most relevant answers; our work does not make such assumptions.

\textbf{(G)STP} has been addressed under \textbf{simplifications} that do not hold in our context.
For instance: the quality of a solution decreases exponentially with the tree size, thus search can stop when all trees are under a certain threshold~\cite{DBLP:conf/edbt/BonaqueCGM16}; edges are considered in a single direction~\cite{DBLP:journals/debu/YuQC10,Elbassuoni:2011:KSO:2063576.2063615,DBLP:journals/tkde/LeLKD14}; the cost function has the suboptimal-structure property~\cite{dpbf,DBLP:conf/sigmod/LiQYM16} etc.
These assumptions reduce the computational cost; in contrast, to leave our options open as to the best score function, we build a feasible solution for the general problem we study.
Some works have focused on finding \textbf{bounded (G)STP approximations}, i.e., (G)STP trees solutions whose cost is at most $f$ times higher than the optimal cost, e.g.,~\cite{soda1998,DBLP:conf/cikm/GubichevN12}.
Beyond the differences between our problem and (G)STP, due notably to the fact that our score is much more general (Section~\ref{sec:score}), non-expert users find it hard to set $f$.

\noindent\textbf{Acknowledgements.} We thank Julien Leblay for his contribution
to earlier versions of this work~\cite{Chanial2018}.
This work has been partially funded by the AI Chair project SourcesSay Grant no ANR-20-CHIA-0015-01 and by Portuguese national funds through FCT with reference UIDB/50021/2020 (INESC-ID).


\renewcommand{\baselinestretch}{0.83}
{\small
\bibliographystyle{abbrv}
\bibliography{references_merged}}
\end{document}